\documentclass[twocolumn]{aa}
\usepackage{graphicx}
\usepackage{txfonts}
\usepackage{xcolor}
\usepackage{hyperref}
\begin{document}
\title{When the tale comes true: multiple populations and wide binaries in the Orion Nebula Cluster}

   \author{Tereza Jerabkova
   \inst{1,2,3}
   \and
   Giacomo Beccari \inst{1}
   \and
   Henri M.J. Boffin\inst{1}
   \and
   Monika G. Petr-Gotzens\inst{1}
   \and 
   Carlo F. Manara \inst{1}%
   \and 
   Pier Giorgio Prada Moroni \inst{4,5}
   \and 
   Emanuele Tognelli \inst{4,5}
   \and 
   Scilla Degl'Innocenti \inst{4,5}
          }
\institute{
European Southern Observatory, Karl-Schwarzschild-Straße 2, 85748 Garching bei München	
\email{tjerabkova@gmail.com}
\and
Helmholtz Institut f\"{u}r Strahlen und Kernphysik, Universit\"{a}t Bonn, Nussallee 14–16, 53115 Bonn, Germany
\and
Astronomical Institute, Charles University in Prague, V 
Hole\v{s}ovi\v{c}k\'ach 2, CZ-180 00 Praha 8, Czech Republic
\and 
Department of Physics "E. Fermi", University of Pisa, Largo Bruno Pontecorvo 3, I-56127 Pisa, Italy
\and 
INFN, Sezione di Pisa, Largo Bruno Pontecorvo 3, I-56127 Pisa, Italy
             }

   \date{Received ; accepted }
\titlerunning{Three stellar populations and wide binaries in the ONC}
\authorrunning{T. Jerabkova and G. Beccari  et al. }

  \abstract
{Recently published high-quality OmegaCAM photometry of the 3x3 deg around the Orion Nebula Cluster (ONC) in $r$, and $i$ filters revealed three well-separated pre-main sequences in the color-magnitude diagram (CMD). 
The objects belonging to the individual sequences are concentrated toward the center of the ONC. 
The authors concluded that there are two competitive scenarios: a population of unresolved binaries and triples with an exotic mass ratio distribution, or three stellar populations with different ages ($\approx  1$ Myr age differences).}
{We use Gaia DR2 in combination with the photometric OmegaCAM catalog to test and confirm the presence of the putative three stellar populations. We also study multiple stellar systems in the ONC for the first time using Gaia DR2.   }
{We selected ONC members based on parallaxes and proper motions and take advantage from OmegaCAM photometry that performs better than Gaia DR2 photometry in crowded regions. We identify two clearly separated sequences with a third suggested by the data. We used Pisa stellar isochrones to estimate ages of the stellar populations with absolute magnitudes computed using Gaia parallaxes on a star by star basis. }
{ 1)  We confirm that the second and third sequence members are more centrally concentrated toward the center of the ONC. In addition we find an indication that the parallax and proper motion distributions are different among the  members of the stellar sequences. The age difference among stellar populations is estimated to be 1-2 Myr. 
2) 
We use Gaia proper motions and other measures to identify and remove as many unresolved multiple system candidates as possible. Nevertheless we are still able to recover two well-separated sequences with evidence for the third one, supporting the existence of the three stellar populations. 3) Due to having ONC members with negligible fore- or background contamination
we were able to identify a substantial number of wide binary objects (separation between 1000-3000 au) and  with relative proper motions of the binary components consistent with zero. This challenges  previously inferred values that suggested no wide binary stars exist in the ONC. Our inferred wide-binary fraction is $\approx 5\%$. }
{We confirm the three populations correspond to three separated episodes of star formation. Based on this result, we conclude that star formation is not  happening in a single burst in this region. In addition we identify 5\% of  wide-binary stars in the ONC that were thought not to be present.}
\keywords{Stars: formation -- Stars: pre-main sequence -- Open clusters and associations: individual: ONC}

\maketitle
%
\section{Introduction}
Young star clusters (YSC) with  resolved  populations are ideal objects for studying and constraining star and star cluster formation. Most of the theoretical and observational works devoted to the study of YSCs seem to agree that their (pre-)stellar population, still embedded in the 
pristine molecular cloud, is mostly coeval with an intrinsic age spread of $\approx 0.5-1$ Myr and a physical size of $\approx 1$ pc~\citep[e.g][]{Zinnecker1993,Lada2003, Pfalzner2011,Marks2012,Getman2018exp}.

However, many authors report on the presence of an age spread up to several Myr in YSCs \citep[][]{Palla2000,Cargile2010,Cignoni2010,Reggiani2011,Bell2013,Balog2016,Getman2018age}. 
Whether the measured age spreads are real and, hence, related to the physics at play during the early stages 
of the formation of the stars in clusters \citep[][]{Larson2003, Pflamm2007, Pflamm2009c,Klassen2017} or related to inaccurate evaluation 
of the impact of observational biases related to differential extinction, stellar variability and complex physical processes like episodic accretion \citep[][]{Baraffe2009,DaRio2010,Jeffries2011} -- is still an open debate. 
 
Recently, \citet[][hereafter B17]{Beccari2017} reported on the detection of three well separated
sequences of pre-main sequence (PMS) stars in an optical color-magnitude diagram (CMD) of an area of $\sim1.5$~deg radius centered on the Orion Nebula Cluster (ONC). The stars belonging to the three
sequences, while being all centered around the Trapezium area, show a different spatial distribution 
with the apparently oldest (and most populous) population being more spatially spread around the center of the ONC with respect to the youngest one which shows an increasing concentration toward the center. 
The authors discussed the possibility that differential extinction and/or a population of unresolved binaries might be at the origin of such an observational feature in their CMD. The effect of differential
reddening is safely excluded by the fact that the three populations show an identical distribution of visual extinction.
While binaries are certainly present as unresolved sources, B17 demonstrated that in order to 
reproduce the separation in the color distribution
of the populations, they require a mass-ratio distribution heavily skewed toward equal mass binaries. While
such a population of binaries could not be safely excluded using the data in hand, B17 stressed that it would
still imply an exotic population of binaries never observed before in a cluster.
In addition, if unresolved binaries or triples were the reason for the discrete sequences in the CMD,  then this would contradict that the binary sequence in this assumption is the most concentrated, because binaries are more easily broken up in denser environments \citep{Kroupa1995b}. 
Supported by recent results from infrared spectroscopy from \cite{dario2016}, they concluded that a real difference in age 
is most likely at the origin of the presence of three populations in the CMD. The populations of PMS stars in the ONC were thus
formed in bursts of star formation separated in age by $\approx 1$~Myr. \footnote{The projection effect (i.e., the emergence of  apparent  unresolved binaries in the inner region due to projection) is larger in the inner region of the ONC and this could conceivably fake a more centrally concentrated younger population. 
The stellar number density at the center is $4.7\times 10^4\,$stars/pc$^3$ \citep{MacCaugherean1994} which implies an average projected separation between two stars at the densest place of about 6000~AU. The resolution of our survey (OmegaCAM and GAIA DR2) is a couple of 100 AU such that the occurrence of an unresolved line-of-sight (non-physical) binary is sufficiently small to be ignored. 
}

\citet[][]{Kroupa2018} proposed a theoretical explanation in support of multiple-bursts of star formation in the ONC, which has been further investigated by \cite{Wang2018}.
In summary, (1) the formation of a first generation, fueled at the hub of inflowing molecular filaments, is truncated
by the feedback of massive stars formed as a part of the first generation; (2) these massive stars are 
stellar-dynamically ejected and as a consequence the inflow of gas still present in
surrounding filaments is restored; (3) the next generation of stars forms. This process 
can be repeated until either the ionizing stars are not ejected or the filamentary gas reservoir is 
exhausted or destroyed. 
The model well predicts the spatial distribution of the three populations as described in B17. 
 
In this paper we use the Gaia DR2 \citep{GAIA_DR2} in combination with OmegaCAM 
photometry to safely isolate the ONC stellar population from fore- or background objects and
to confirm or discard the detection of multiple sequences in the optical CMD. The impact of
unresolved binaries will be studied in detail.
If the presence of multiple populations in the ONC is confirmed, the case of the ONC may 
represent a fruitful testing site for theoretical models of star cluster formation including 
the potentially important process of its stellar-dynamical modulation. 
In addition we take advantage of 
the Gaia spatial resolving power and analyze apparent multiple system candidates using isolated (that is, with negligible fore- or background contamination) ONC star members only. 
Such a study is possible for the first time thanks to Gaia.

The manuscript is structured as follows: In Sect~\ref{Sec:Data} we introduce our data sets, 
then we apply selection criteria to obtain ONC members
(Sect~\ref{Sec:ONC}). Using this sample we recover the sequences found by B17 and discuss their properties in Sect~\ref{sec:seq}. The study of the binary stars allowed by Gaia is described in Sect~\ref{sec:Rbin}, followed by a discussion and the conclusions in Sect~\ref{Sec:Conc}.

\section{Data sets} \label{Sec:Data}

\subsection{The Gaia data}
We used the python Astroquery package 
\citep{gaia_py} to retrieve the Gaia DR2 data \citep{GAIA_DR2} from the Gaia science archive\footnote{http://gea.esac.esa.int/archive/}. 
We downloaded all the objects detected by Gaia that are within a radius of three degrees on the sky from the ONC 
(R.A.$ \approx 83.75 \, \mathrm{deg.}$, Dec $ \approx -5.48 \, \mathrm{deg.}$, 1 deg on the sky corresponds to $\approx$ 7 pc at the ONC distance of 400 pc) without any additional filtering. 
The data sample contains 278,444 targets, of which $12\%$ have parallaxes measured with a relative error of 10\% or smaller, while $15 \%$ do not have a measured parallax. The data sample is dominated by faint stars, with only $43\%$ stars being brighter than 19 mag in the Gaia G-band. 

We note that Gaia DR2 photometry in BP and RP filters is not suitable to study the dense and still in-gas embedded regions as the ONC. We follow the Gaia DR2 quality data check in the Appendix \ref{sec:filt} and Sect~\ref{sec:bin} to study Gaia DR2 photometric capabilities on our data set and conclude that for our analysis it is necessary to use additional photometric data. 

\subsection{The OmegaCAM catalog}
\label{sec:data_sample}
We acquired a new set of deep multi-exposures with OmegaCAM, attached to the 2.6-m VST telescope in Paranal, with the aim of increasing the photometric 
accuracy toward the faintest region of the optical CMD shown in B17. We surveyed a $3\times3$ deg area around 
the center of the ONC using the $r$ and $i$ filters, that is, the same bands as adopted in B17. In particular 
we acquired $10\times25$s exposures in both filters for each pointing under the proposal 098.C-0850(A), PI Beccari. As in B17, the whole data reduction process (from removal of detector signatures to the extraction and calibration 
of stellar magnitudes) was carried out at the Cambridge Astronomical Survey Unit (CASU). We downloaded 
the astrometrically and photometrically calibrated single band 
catalogs from the VST archive at CASU\footnote{http://casu.ast.cam.ac.uk/vstsp/}. We used a large number of 
stars in common with the AAVSO Photometric All-Sky Survey (APASS) to correct for possible residual offsets in 
$r$ and $i$ magnitudes between each single exposures. We finally require that a single object should be detected in 
at least seven out of ten images both in the $r$ and $i$ band in order to be accepted as a real star. Since the error on individual stellar sources is very small (see Fig.~\ref{fig:CMD_sum}), the average of 
the magnitudes measured in each individual frame was adopted as stellar magnitude while the standard  error 
was used as the associated photometric error. The final catalog includes 93846 objects homogeneously sampled 
in $r$ and $i$ bands down to $r \approx 21-22 \, \mathrm{mag}$ on an $3\times3~\mathrm{deg}$ area around the cluster's center.

\subsection{The initial catalog} \label{sec:incat}
We used the $C^3$ tool from \cite{Riccio2017} in order to identify the stars in common between the OmegaCAM and the Gaia catalog. $C^3$ is a command-line open-source Python script that, among several other options, can cross-match two catalogs based on the sky positions of the sources.  We used the $C^3$ {\it best} matching option and circular selection region with a radius of 50 arcsec that is used for sky partition.
We found 84022 targets in common between the Gaia and the OmegaCAM data. The majority of targets that are in the OmegaCAM catalog but not in the Gaia one are faint ($r\gtrapprox 21$) and blue objects ($r-i \lessapprox 1.2$). We note that we also loose a fraction of objects in the most crowded regions, like the center of the ONC (see Appendix \ref{sec:filt}).
Hereafter we refer to the Gaia DR2 $and$ OmegaCAM cross-matched catalog as the initial catalog.

\section{The ONC members with Gaia} \label{Sec:ONC}
As described in the previous section, our initial catalog includes optical photometry from OmegaCAM and 
astrometric informations from Gaia DR2  for objects in the range
of mag $12\lesssim r\lesssim20$ within $3\times3$ deg around the center of the ONC. We now propose a roadmap which allows users to fully exploit the potential offered by
the Gaia DR2 astrometric information in efficiently disentangling the stellar populations of a cluster
from field objects in the CMD. Such a strategy is  adopted here to specifically identify and study the PMS population(s) in the ONC, but we are confident that such criteria can be applied to perform similar studies of stellar populations in close-by young clusters.

\begin{figure*}[ht!]
 \scalebox{1.0}{\includegraphics{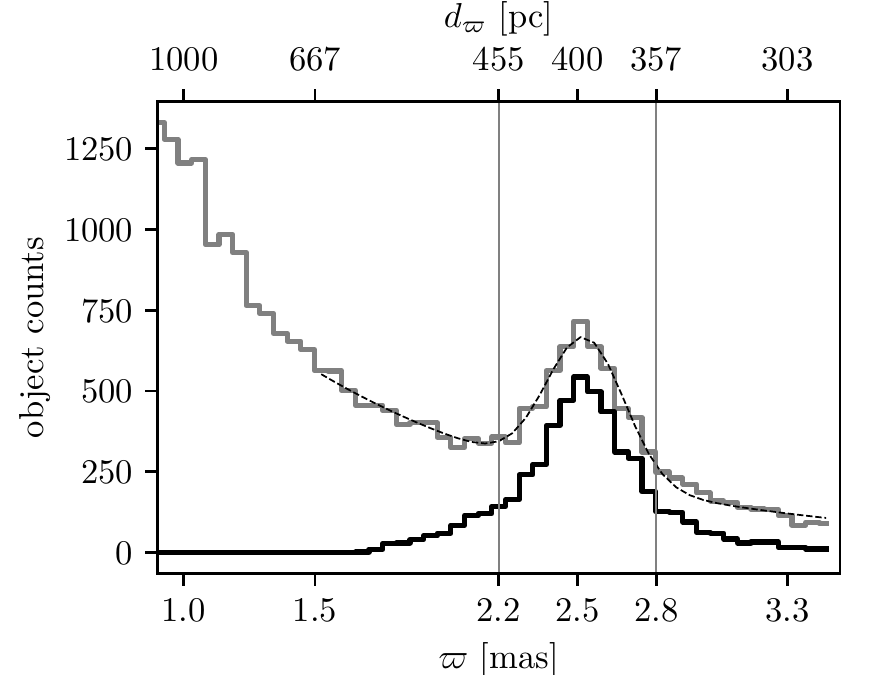}}
 \scalebox{1.0}{\includegraphics{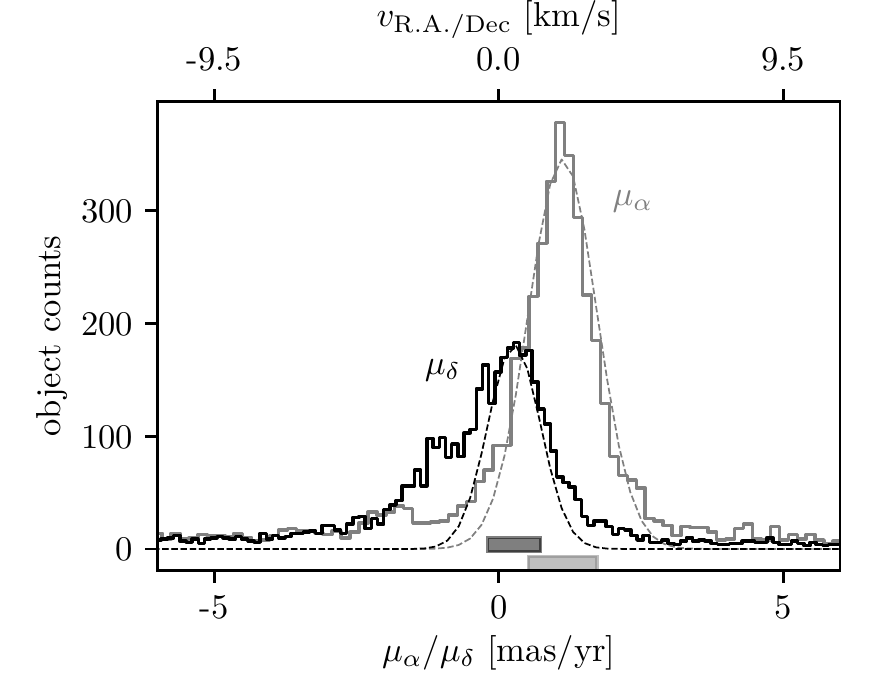}}
  \caption{\textbf{Left panel:} distribution of the parallax, $\varpi$, of all the objects in the initial catalog (gray histogram) and the bona fide ONC members selected with C1 (black histogram; see Sect.~\ref{Sec:ONC}). 
The dashed line show the Gaussian fit to the distribution of parallaxes of the stars in the initial catalog. The vertical
line indicates the $\pm 2\sigma$ distance from the peak of the distribution at $\approx 2.52$.
The top horizontal axis shows corresponding values, $d$, in parsecs using $d \mathrm{[pc]}= 1/(\varpi \mathrm{[mas]} \cdot 10^{-3})$, valid for those objects that have small relative errors \citep[][]{GAIA_DR2}.  
\textbf{Right panel:} Distributions of the  R.A. and Dec proper motions (gray and black histograms, respectively). 
To select the objects that have proper motions consistent with the bulk of the ONC we fit each distribution with a Gaussian function shown by dashed lines. The gray horizontal bars indicate the $1\sigma$ range of the respective fit. The top axis shows the values of the proper motions in km/s, calculated assuming a distance of 400 pc.}
  \label{fig:par}
\end{figure*}

\subsection{Constraining the reliable ONC members: Parallax selection - C1}
 
We first use the parallaxes provided by Gaia DR2 to reliably identify ONC members. 
We show as gray histogram in Fig.~\ref{fig:par} the distribution of the parallaxes $\varpi$ for all 
the stars belonging to our initial catalog. 
The stellar population belonging to the ONC is easily recognized through a well defined peak at $\varpi_{\mathrm{peak}}\approx 2.5$ mas corresponding to a distance $d \approx 400$~pc, in full agreement with previous estimates of the ONC distance~\citep[e.g.,][]{me07}. 
This plot already witnesses
the unprecedented opportunity offered by the Gaia-DR2 catalog to safely identify and hence study the details of
the stellar populations in clusters. 

In order to isolate the ONC members, we performed a fit to the parallax distribution (dashed line in Fig.~\ref{fig:par}) using a Gaussian function in combination with an exponential function. The latter is appropriate to describe the fore- or background sources. The fit shows a peak at 
\begin{eqnarray*}
\varpi_{\mathrm{ONC}} = 2.52 \pm 0.15\,\mathrm{mas}\, ,
\end{eqnarray*}
where the uncertainty is  1-$\sigma$ of the Gaussian fit. Note that with $\sigma$ we here indicate the uncertainty related to the Gaussian fit while elsewhere in the paper we indicate as $\sigma_{\varpi}$ the error associated to the value of ${\varpi}$ as given in the Gaia catalog.
To ensure that we only keep the best candidates, 
we retain as bona fide members all the objects whose parallax falls
in the range defined by $\pm 2\sigma$  distance from the peak of the distribution (that is  2.22 and 2.82~mas -- the vertical lines in Fig.~\ref{fig:par}). In addition, only the targets having a relative parallax error within 
10\% are considered as this ensures good quality of the astrometric solution~\citep[see also][]{GAIA_DR2}.
The final selection criterion C1 is purely based on the parallaxes and is defined as:
\begin{equation}\label{eq:C1}
2.22 - 3\sigma_{\varpi} \leq \varpi \leq 2.82 + 3\sigma_{\varpi} \,  \texttt{and} \, \sigma_{\varpi}/\varpi \leq 0.1 \, ,
\end{equation}
where $\sigma_{\varpi}$ is the uncertainty on each single parallax given in Gaia DR2. 

The black solid line in Fig.~\ref{fig:par} shows the distribution of the
parallaxes of the 4988 stars selected using the parallax filtering C1 (Eq.~\ref{eq:C1}). 
We include the stellar population
in a $\approx 100\,\mathrm{pc}$ region from the ONC along the line of sight. Such a region is somehow larger than the physical extension 
of the ONC star forming region on the sky (e.g., B17 or our Fig.~\ref{fig:bin_sky}).  We note, however, that having 10\% uncertainty in parallax at the distance of the ONC results in  $\approx$ 40 pc uncertainty on the distance inferred from the parallax. 

By adopting the criterion C1 (Eq.~\ref{eq:C1}), we uniquely rely on the use of parallaxes, 
implying that targets with available optical photometry from OmegaCAM but not having precise enough cataloged parallaxes are not considered. 
Such objects are usually faint or are residing in crowded fields, like the ONC center. 
In addition multiple or close by objects for which a single star's astrometric solution was not found would also lack 
cataloged astrometric parameters.
We stress here that 12\% of the stars in the initial photometric catalog do not have parallaxes. This fraction is reduced to 
4\% if we only consider objects that are in the CMD position occupied by the young stellar population of the ONC.

\subsection{Constraining the reliable ONC members: Proper motion selection - C2}

As previously mentioned, by adopting the C1(Eq.~\ref{eq:C1}) criterion we isolate the stars in a region with a depth of $\approx 100$ pc around the ONC. 
Since this area is much larger than the expected physical size of the ONC \citep{Hillenbrand1997}, it is not unexpected that the use of the C1(Eq.~\ref{eq:C1}) criterion alone does not fully filter out fore- or back-ground objects. 
Since the stars belonging to a given cluster do in general show similar values of proper motions ($\mu_{\alpha, \delta}$), 
we can use the $\mu_{\alpha, \delta}$ available from Gaia DR2 to further constrain the ONC membership of the stars selected with the C1(Eq.~\ref{eq:C1}) criterion. 
We show in the right panel of Fig.~\ref{fig:par} the distributions of $\mu_{\alpha}$ and $\mu_{\delta}$ (gray and black histograms, respectively)
of all the stars in the C1(Eq.~\ref{eq:C1}) selected data set.

It is interesting to note that the distributions of $\mu_{\alpha, \delta}$ are not Gaussian. The histograms show a prominent main peak together with one or more sub-peaks populating the wings of the distributions. 
Such sub-peaks might be indicative of a system of sub-clusters or streams located in the vicinity of the main cluster and likely born out of the same molecular cloud. 
We will investigate this interesting possibility and the detailed nature of such sub-peaks in a forthcoming work, while here we only focus on the main population of the ONC star cluster. 
To further isolate the bona fide ONC members we hence fit a Gaussian function to the main peaks only. 
The two Gaussian functions are shown as dashed lines in the right panel of Fig.~\ref{fig:par} with peaks at
\begin{equation}\label{eq:ONCpm}
\begin{aligned}
\mu^{\mathrm{ONC}}_{\alpha} &=& 1.1\pm0.6\, \mathrm{mas/yr},\\ 
\mu^{\mathrm{ONC}}_{\delta} &=& 0.3\pm0.5\, \mathrm{mas/yr}. 
\end{aligned}
\end{equation}
These values are consistent with the ones derived by \cite{Kuhn2018} 
 for the inner most 378 selected members, $\mu^K_{\alpha \, \star} = 1.51 \pm 0.11\,\mathrm{mas\, yr^{-1}}$, $\mu^K_{\delta} = 0.5\pm0.12\,\mathrm{mas\, yr^{-1}}$.
The selection criterion $n\sigma$ C2 is defined in the PM space and allows us to include only the stars whose PM in R.A. and in Dec fall inside a $n\sigma$ range as follows,
\begin{equation}\label{eq:C2}
\begin{aligned}
\mu^{\mathrm{ONC}}_{\alpha} - n\sigma_{\mathrm{F\,\mu_{\alpha}}} &\leq \mu_{\alpha} \leq  + \mu^{\mathrm{ONC}}_{\alpha} + n\sigma_{\mathrm{F\,\mu_{\alpha}}}\, ,\\
\mu^{\mathrm{ONC}}_{\delta} - n\sigma_{\mathrm{F\,\mu_{\delta}}} &\leq \mu_{\delta} \leq  + \mu^{\mathrm{ONC}}_{\delta} + n\sigma_{\mathrm{F\,\mu_{\delta}}}\, ,
\end{aligned}
\end{equation}
where $n\sigma_{\mathrm{F\,\mu_{\alpha}}}$ and $n\sigma_{\mathrm{F\,\mu_{\delta}}}$ are the number of $\sigma$ values of the Gaussian to the $\mu_{\alpha,\delta}$ distributions, respectively. The 1-$\sigma$ values  denoted as $\sigma_{\mathrm{F\,\mu_{\alpha}}}$ and $\sigma_{\mathrm{F\,\mu_{\delta}}}$ are shown in the right panel of Fig.~\ref{fig:par} as horizontal light and dark gray bars. Their estimated values are written in Eq.~(\ref{eq:ONCpm}).
While the combined use of the C1(Eq.~\ref{eq:C1}) and C2(Eq.~\ref{eq:C2}) criteria is quite powerful in removing fore- or backgrounds objects, it certainly also removes genuine ONC members. Still we stress here that such objects show PM properties that deviate from the main ONC populations. Such objects are likely binary systems with separations $\approx 0.1"-0.4"$  
or stars sitting in regions affected by severe crowding where the performance of Gaia degrades and the astrometric solution can be spurious \citep{Arenou2018,Lindergren2018}. 
In the first case, the use of C1~(Eq.~\ref{eq:C1}), and even more so C2~(Eq.~\ref{eq:C2}), very nicely
serves one of the main goals of this study to verify the presence of multiple
populations in the ONC reported by B17 (see Sec~\ref{sec:bin}). In the second case, removing genuine ONC stars that are mostly sitting in the
center of the ONC where most of the stars belonging to the youngest population are found, might delete statistically significant kinematic 
signatures. 
Hence, the rigorous selection criteria adopted in this work might potentially weaken or even remove any signature of multiple sequences in the CMD. On the other hand, we are confident
that, by applying these filters, if any signature of multiple populations is still observed, this would be a solid observational support toward the presence of multiple and sequential bursts of star formation in the ONC. 

\begin{figure*}[htbp]
  \scalebox{1.0}{\includegraphics{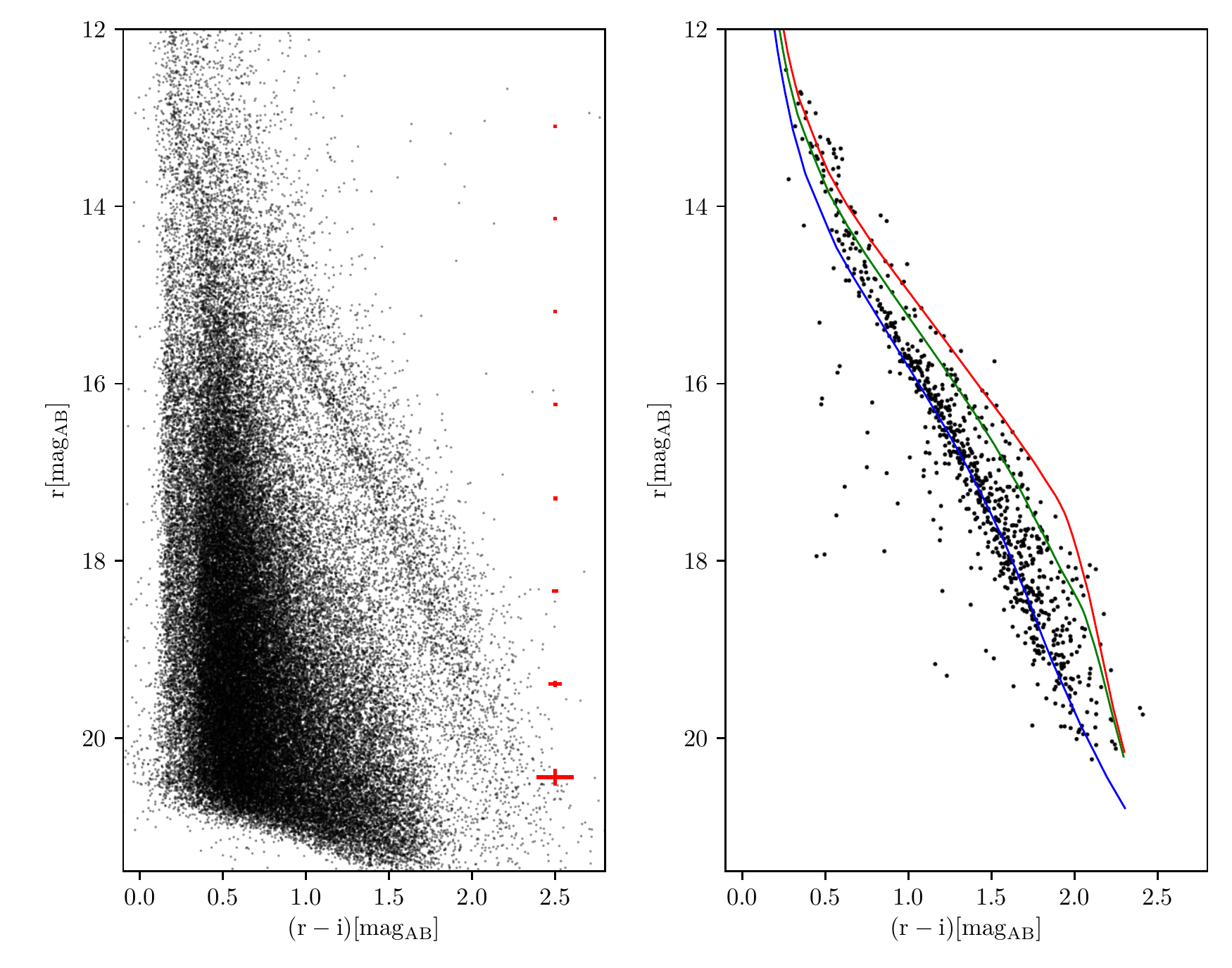}}
  \caption{{\bf Left panel:} CMD for the initial catalog. The red crosses show $3-\sigma$ color and magnitudes errors. {\bf Right panel:} The CMD of the parallax and proper motion selected ONC members (criteria C1(Eq.~\ref{eq:C1}) and 1$\sigma$ C2(Eq.~\ref{eq:C2}) and C3). The color lines are the best fitting isochrones from the Pisa stellar evolutionary model of the three populations.
}
  \label{fig:CMD_sum}
\end{figure*}

\subsection{Constraining the reliable ONC members: Gaia quality check - C3} \label{sec:bin}
There are several factors that can affect the accuracy of the
astrometric and photometric parameters measured by the Gaia satellite 
\citep{Lindergren2018,Evans2018,Arenou2018}. 
Among others, stellar crowding in a high density environment, faintness of the stars and stellar multiplicity are of particular interest for our work. 
Such effects are at the origin of the fact that for a number of objects observed with OmegaCAM, parallaxes were not available or
had too large uncertainty (C1; see above). The Gaia catalog offers several  parameters that can be used to verify the quality of the measures associated to given objects~\citep{Arenou2018}. In the following we describe a few parameters that are of particular interest for our work:

\begin{itemize}
\item Duplicated source criterion:\\ 
The current version of the Gaia catalog cannot provide any reliable measurements for sources whose separation on the sky is lower than $0.4-0.5\arcsec$~\citep{Lindergren2018,Arenou2018}.
Such objects are flagged in the Gaia-DR2 as duplicated sources (\texttt{duplicated\_source == True}).
Hence, it is immediately understandable that this parameter can be used to identify objects whose measurements are affected by high stellar density (crowding) or unresolved multiplicity. 
According to \citet{Arenou2018} the average separation among duplicated targets is 0.019\arcsec, corresponding to $\approx 8\,$au at the distance of the ONC ($\approx 400\,$pc). In summary, when we apply the criterion C3 we remove all the stars which appear with the \texttt{duplicated\_source == True} in the Gaia catalog.

\item Astrometric: \\
 \citet{Arenou2018} defined the following criterion: 
\begin{eqnarray}
u \leq  1.2 \max{(1,\exp{(-0.2(G-19.5)))}}\, ,
\label{eq:ucrit}
\end{eqnarray}
where $u = \sqrt{\chi^2/\nu}$, 
$\chi^2 =$ \texttt{astrometric\_chi2\_al} and
$\nu =$ (\texttt{astrometric\_n\_good\_obs\_al }$-5$) 
suggesting that it can be used to check for spurious astrometric solutions. 
In the first panel of Fig.~\ref{fig:uE}, the $u$ parameter is shown as a function
of the apparent Gaia $G$ magnitude. The red dashed line shows Eq.~(\ref{eq:ucrit})
for the equal sign. Following the study of \citet{Arenou2018} we will reject from our study all objects whose value of $u$ falls above this line. 
\end{itemize}

The Duplicate source criterion together with the Astrometric criterion are called the C3 criterion.
We show in the left panel of Fig.~\ref{fig:CMD_sum} the CMD of all the stars included in our initial catalog. In the right panel of the same figure
we instead show the CMD of the 852 bona fide ONC stars selected using the C1 (Eq.~\ref{eq:C1}) and $1\sigma$~C2 (Eq.~\ref{eq:C2}) and C3 filtering criteria described above. The population of PMS stars belonging to the ONC is identified with unprecedented accuracy in our final catalog. 
A main population of PMS is clearly detected while a second parallel sequence is also clearly visible. We emphasize here that
using the criteria C1+C2+C3 we already removed many unresolved binary systems.

As previously mentioned, the potential biases introduced by the filtering certainly affect the star counts in the very central part of the cluster because of the presence of stellar crowding and multiple systems. In the next section we will use the catalog of bona fide members to quantitatively investigate  the presence of multiple sequences as reported in B17, and their relation to multiplicity. 
We stress here that the youngest (reddest) populations detected in B17 are also more concentrated toward the center of the cluster. 
Therefore the biases introduced by the C1(Eq.~\ref{eq:C1}), C2(Eq.~\ref{eq:C2}) and C3 criteria will certainly affect the statistical significance of the detection of the redder (and possibly younger) sub-populations.
\\

\section{Identification of multiple sequences in the CMD}\label{sec:seq}

\begin{figure*}[htbp]
  \scalebox{1.0}{\includegraphics{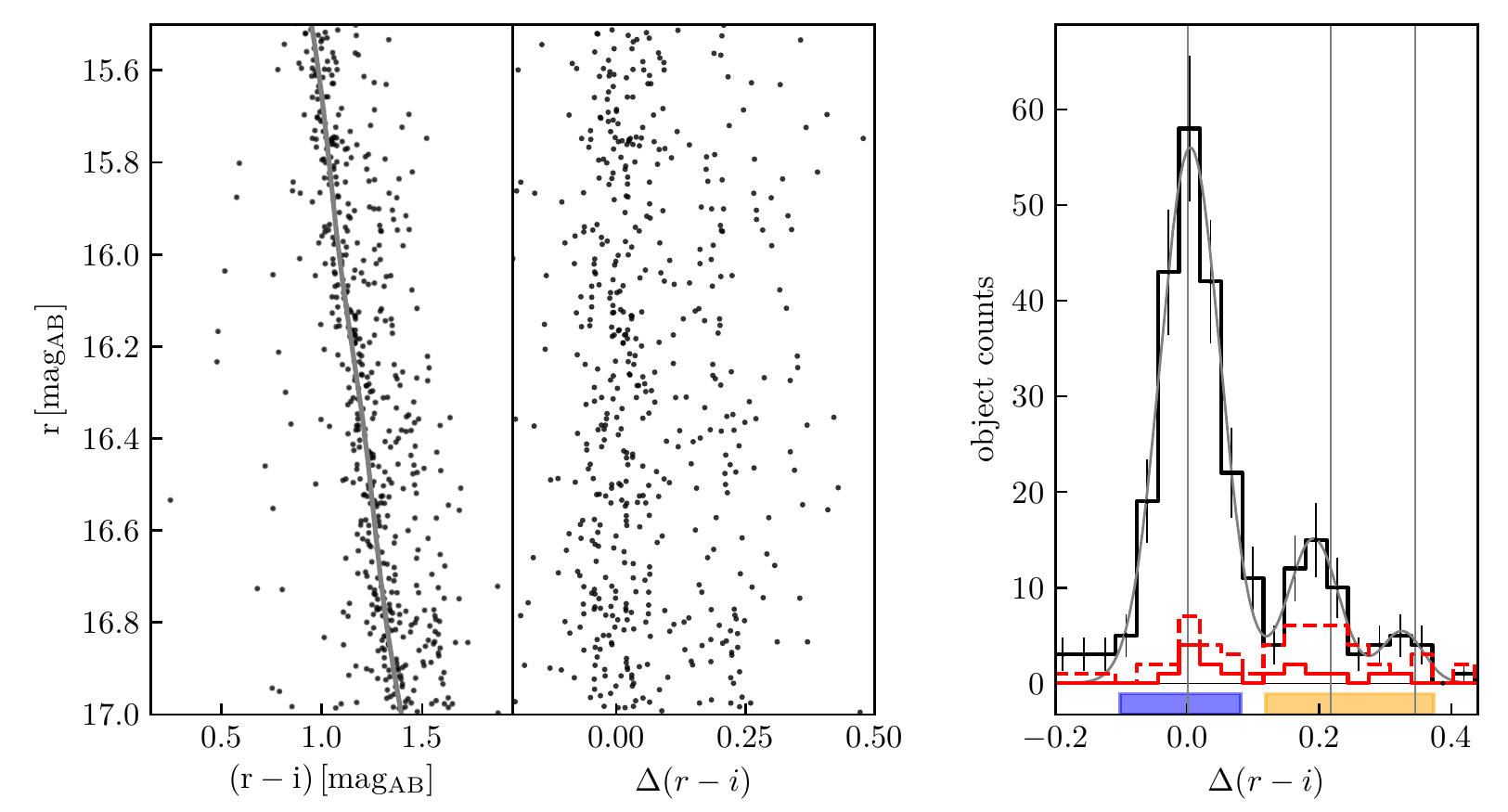}}
  \caption{
{\bf Left panel:} CMD of the C1+C2-1$\sigma$+C3 selected data sample. The main ridge line is shown as a gray line.
{\bf Middle panel:} The CMD rotated along the slope of the main ridge line.	
{\bf  Right panel:} The histograms of the distances in color from the main ridge line. The black solid line corresponds to the data sample after applying all selection criteria (C1, C2 1-$\sigma$ and C3).
The vertical line at $\Delta (r-i)=0$ shows the position of the mean ridge line which serves as a reference line. It also represent the location of the single stars belonging to the main population of the ONC. The two vertical lines $\Delta (r-i)>0$ indicate the expected positions of the equal mass binary and triple systems, respectively. The red dashed histogram shows the unresolved binary stars from \cite{Tobin2009}. The solid-line red histogram shows the distribution of the unresolved binaries that passed the selection criteria (C1, C2 1-$\sigma$ and C3).}
  \label{fig:CMD_sell}
\end{figure*}

\begin{figure}[ht]
  \scalebox{0.9}{\includegraphics{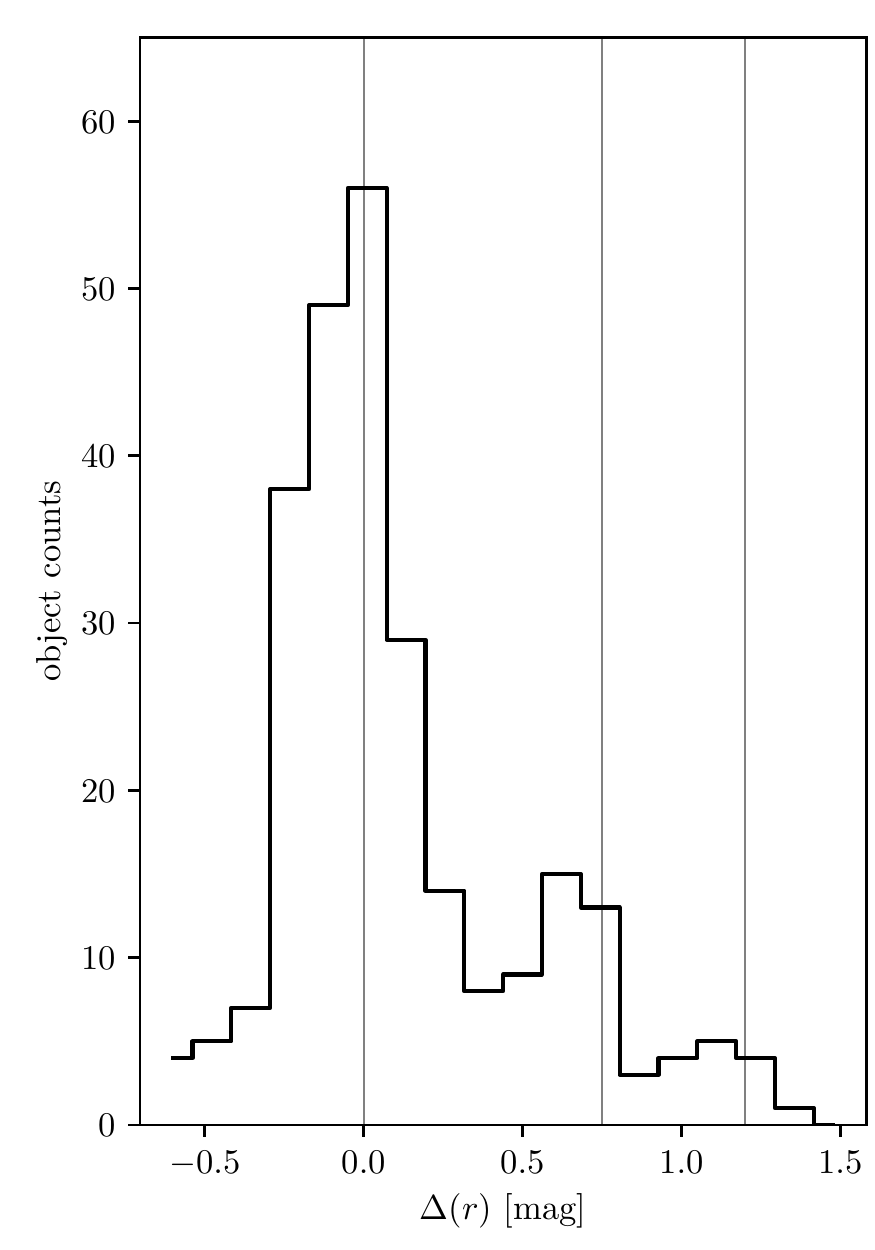}}
  \caption{ Complementary plot to Fig.~\ref{fig:CMD_sell} showing the distribution of magnitudes from the main ridge line instead of the perpendicular distance in color and magnitude that is on Fig.~\ref{fig:CMD_sell}. We show the position of the main ridge line (at 0.0) as well as unresolved equal-mass binaries and equal-mass triples stars, using vertical lines. The positions of the three peaks relative to the main ridge line and equal-mass binaries or triples is quantitatively comparable to the  Fig.~\ref{fig:CMD_sell}.
  }
  \label{fig:ref}
\end{figure}

In order to identify the presence of multiple and parallel sequences of PMS objects in the ONC, we applied the same approach as described in B17. First, we calculate the main ridge line of the bluest and most populated PMS population in the range of magnitudes 15.5 < $r$ < 17 (left panel of Fig.~\ref{fig:CMD_sell}). We use the mean ridge line as reference in the ($r, r-i$) CMD and we then calculate the projected distance in $r-i$ colors of each star from the reference line (see the central panel of  Fig.~\ref{fig:CMD_sell}). We show in the right panel of Fig.~\ref{fig:CMD_sell} the distribution of the perpendicular distances in color  from the main ridge line, $\Delta(r-i)$.
The black histogram shows the distribution of C1+1$\sigma$C2+C3 selected data sample.
To complement Fig.~\ref{fig:CMD_sell} that is showing projected distribution $\Delta(r-i)$ from the main ridge we show, on the request of the anonymous referee, a distribution in vertical distances from the main ridge line $\Delta(r)$, see Fig.\ref{fig:ref}.

Clearly the distribution in color of the ONC members is described
by at least two peaks with a gap in the middle. This feature fully confirms the observation in B17. As already noted by B17, the number of stars belonging to the third population (i.e., the reddest peak) is quite low. We would like to stress here once more that the use of the C1(Eq.~\ref{eq:C1}), C2(Eq.~\ref{eq:C2}) and C3 filtering preferentially removes stars located in crowded regions. We hence expect that our selection criteria mostly lower the number of stars
in the reddest or youngest population that, according to B17 is the one that is most concentrated toward the ONC center.

We used three Gaussian functions (gray line in the figure) to fit the color distribution and in particular the three peaks. We then use these functions to select members belonging to each population. Given the fact that the second and third sub-populations are overall populated by a very low number of stars (54 and 15, respectively) we decided to proceed further with our analysis by dividing the ONC populations into an "old" and "young" population (respectively blue and orange bar in Fig.~\ref{fig:CMD_sell} and afterwards). Hence the stars belonging to the second and third peak are grouped and studied together.

\begin{figure}[ht!]
  \scalebox{1.0}{\includegraphics{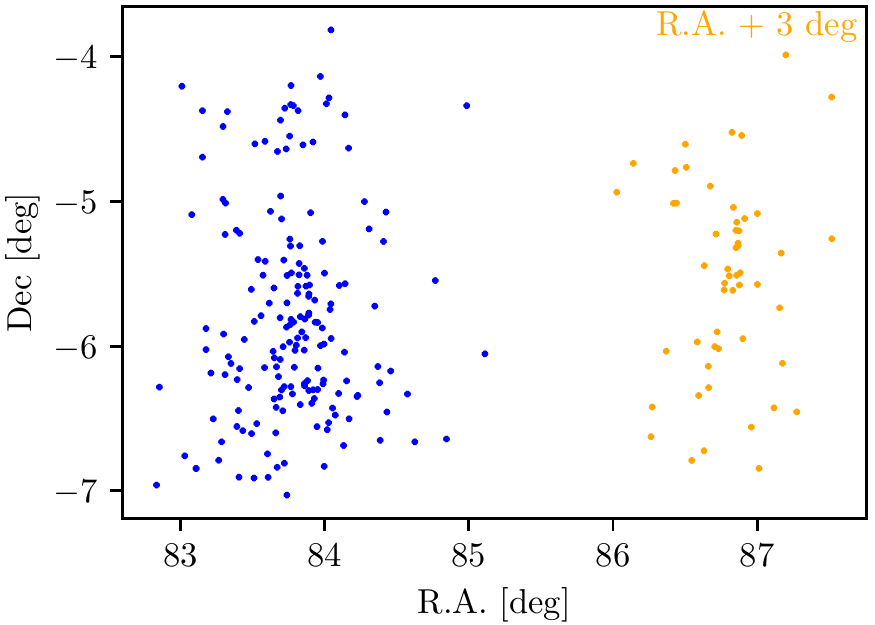}}
  \caption{The R.A. - Dec distribution of the first and second sequence identified in Fig.~\ref{fig:CMD_sell}. 
  To compare on-the-sky distributions of the "old" and "young" sequences we use two-dimensional Kolmogorov-Smirnov test \citep{Peacock1983,Fasano1987,Press2007} resulting in $p=0.01\,(\approx 2.3\sigma)$. 
  } 
  \label{fig:radec}
\end{figure}

In Fig.~\ref{fig:radec} we show the distribution on the sky of the two sub-populations (blue and orange points). The two-dimensional Kolmogorov-Smirnov test \citep{Peacock1983,Fasano1987,Press2007} done on the measured distribution suggests that the two populations are not extracted from the same parental population at $\approx 2.3 \sigma$ level of significance. 
 We would like to stress here that the data set shown in the figure suffers from severe incompleteness, especially in the central region, because of the Gaia DR2 loss of sensitivity in regions affected by stellar crowding. As extensively discussed by B17, the use of the OmegaCAM data-set alone (i.e., without applying the filtering used in this work) indicate the presence of a prominent concentration of all three populations toward he same center.

\begin{figure*}[ht!]
 \scalebox{1.0}{\includegraphics{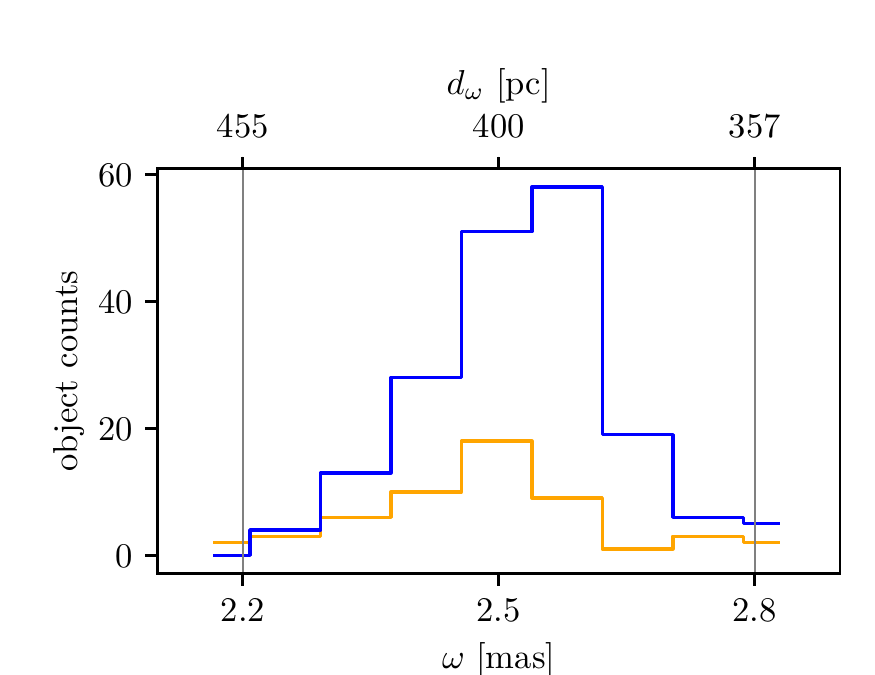}}
 \scalebox{1.0}{\includegraphics{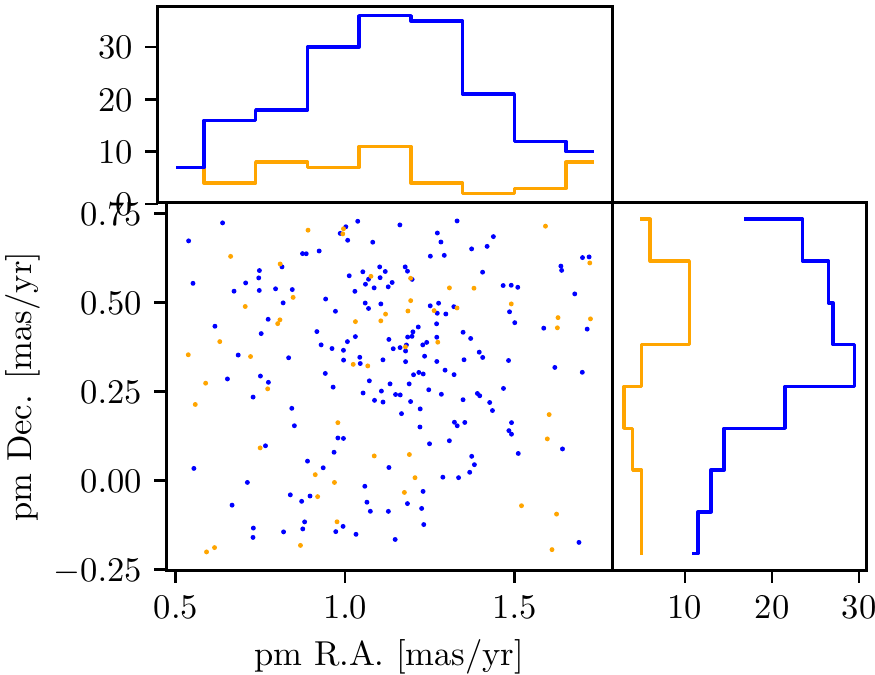}}
  \caption{
Left panel: distribution of the parallax, $\varpi$, of the identified main-ridge sequence (blue), second +third sequence (orange). 
The top horizontal axis shows corresponding values, $d$, in parsec using $d \mathrm{[pc]}= 1/(\varpi \mathrm{[mas]} \cdot 10^{-3})$, valid for those objects that have small relative errors \citep[][]{GAIA_DR2}.  
Right panel: the  R.A. and Dec proper motions with the histograms of projected distributions on the sides of the plot.}
  \label{fig:pop}
\end{figure*}

Gaia offers us the possibility to study, for the first time, the parallax distribution and the proper motions of the two sub-populations. We plot in Fig.~\ref{fig:pop} the histogram of the parallax (left panel) and proper motions (right panel) distributions of the blue and orange populations. The 2D KS test indicates that the proper motions of the blue and orange sub-populations are not extracted from the same parental population at only $\approx 1.3\sigma$. For the parallax distributions, which are 1D, the Anderson-Darling statistical test~\footnote{implemented in the \texttt{scipy.stats} python package based on \cite{Anderson_test}} indicates that they are not extracted from the same parental distribution at $\approx 2.3\sigma$ \citep[see][ for suggested line-of-sight complex distribution of stellar populations]{Alves2012}.
Since the members of the ONC were selected such that they have the same proper motion values (see Sect \ref{sec:data_sample}) the potential difference in proper motions for the different populations might have been erased. However, it is the use of proper motions that offers the most reliable way to constrain the membership in the ONC with Gaia DR2 as the parallax uncertainties are still too large. The 2.3$\sigma$ difference is thus suggestive but not conclusive evidence of a real difference in proper motions and needs to be investigated further with future releases of the Gaia catalog.

We used the Anderson-Darling test to explore the possibility that the comparison of the distributions of proper motions are not affected by different distributions in magnitudes among the different stellar populations. The test indicates that the distributions of magnitudes of the bluest population and the two reddest ones combined are extracted from parental populations which cannot be distinguished significantly ($p=0.4$ for the magnitudes not to being drawn from the same parental distribution).

To summarize, the currently available data indicate that the ONC objects are gradually more centrally concentrated (proportionally to $\Delta (r-i)$) and suggest potential differences in the parallax distributions between the sequences. Under the assumption that the bluer population is older with respect to the orange one (see B17) the current analysis indicate that the older population of the ONC might be closer to us with respect to the younger one.
Both of these findings seems to support the hypothesis that a genuine difference in age is at the origin of the distribution in the CMD of the two sub-populations.

\subsection{Multiple sequences in the ONC with binary star analysis}

Here we would like to study the presence of unresolved multiple systems among the 
catalog used to obtain the histograms of Fig.~\ref{fig:CMD_sell}. 
Ideally, the main ridge line of the main population (corresponding to $\Delta(r-i) = 0$)
represents the position of single stars. In reality single stars will have certain scatter due to intrinsic photometric uncertainty and possible age spreads. 

Unresolved binary system will have a total magnitude in the range of the primary component up to being brighter by 0.75mag in the case of equal mass binaries. Thus the upper theoretical limit for $\Delta (r-i)$ of unresolved binaries is constructed by projecting the theoretical location on the CMD of the equal mass binary sequence along the main ridge line, as done for the observed stars. 
The same exercise can be done for the equal mass triple systems. The two  gray vertical lines in the
right panel of  Fig.~\ref{fig:CMD_sell} represent the theoretical maximum values in $\Delta (r-i)$ of any unresolved binary and triple systems. 
But this is not realistically the case because the second and third peaks are too defined, since an age dispersion would soften the peaks if they were due to unresolved binaries and triples. That is, in order to explain the outlying systems through a small age spread, one would need an even more unnatural mass-ratio distribution than already excluded by B17.
While it is possible that the objects falling in between the reference line at $\Delta (r-i)=0$ and the one indicating the position of multiples are indeed unresolved binaries, one should invoke a peculiar mass ratio distribution in order to reproduce the observed color distribution and the presence of a gap (see B17 for explicit calculations).
The separation of the main ridge line and the second peak remain the same as in B17 and thus we do not repeat the calculations here. We would need to do exactly the same as in B17 with those same data, since the data used here are subject to, among other biases as stated above, the central incompleteness that certainly affects the shape of the distributions in $\Delta (r-i)$.

\cite{Tobin2009} published a catalog of 135 spectroscopic binaries (SBs) in the ONC. All the SBs are recovered 
in our initial catalog.  
We show as a gray dashed histogram in Fig.~\ref{fig:CMD_sell} the distribution in $\Delta (r-i)$ of the unresolved binary stars from \cite{Tobin2009} falling in the magnitude range 15.5<r<17 (64 out of 135). Among the 64 unresolved binaries, 30 populate the $\Delta (r-i)$ range coincident with the bluest and oldest population of the ONC, while 34 are in the reddest peak. This sample allow us to test how efficient is our filtering methods in identifying unresolved multiple systems.

In fact we applied to the catalog of 64 unresolved binaries from \cite{Tobin2009} and retrieved in our photometry the C1, C2 and C3 selection criteria. The distribution in the $\Delta (r-i)$ plane of the catalog of filtered binaries that survive to our selection criteria is shown as solid  red line in the right panel of Fig.~\ref{fig:CMD_sell}. Only 16 stars out of the initial 64 (25\%) passed our filtering. In particular, nine stars among the 30 objects populating the bluest (30\%) peak remains undetected, while only seven among the reddest 34 systems passed our selection criteria (20\%). 
This test demonstrates that our filtering method largely removes the unresolved binary stars, especially those that are in the position of the second peak (that is $q\gtrapprox 0.5$). So it indeed shows that our filtering method is able to remove binaries that are present in the second sequence. 
To confirm this we performed the following experiment: we define as the general population, the sample that has been parallax selected with a 3-sigma cut in proper motion, that is, we keep only the likely members of the regions (noting that such a sample is incomplete). For this general population the above-applied filtering keeps 60\% of the stars belonging to the main peak (population one) and 40\% of the stars belonging to the second and third peak -- proving a) that the filtered number of targets
is smaller for the general population data sample than for the binary star catalog by \citet{Tobin2009}, in agreement with the statement that the applied filtering is able to remove binary stars, and b) that the removed fraction of stars in the range where we expect unresolved binary or triple stars is higher, in line with the previous point a).

We note that the filtering removes a substantial fraction of multiple systems. In particular, Gaia cannot find astrometric solutions typically for binaries. The exact fraction of binaries filtered out by which of the criteria is however a non-trivial problem and cannot be quantified here as this would depend on detailed modeling of the space craft and the stellar population at the distance of the ONC, see for example \cite{Michalik2014}.

\subsection{PMS isochrones}
Stellar models have been computed using the most recent version of the Pisa stellar evolutionary code \citep[see e.g.,][]{tognelli2011} described in detail in previous papers \citep[see][and references therein]{randich2018,tognelli2018}.
We simply recall the main aspects relevant for the present analysis. The reference set of models adopts a solar calibrated\footnote{Solar calibration is an iterative procedure to derive the initial metallicity, helium content and mixing length parameter needed to reproduce, at the age of the Sun, for a 1 $M_{\odot}$ model the Solar luminosity, radius and surface $(Z/X)$ composition, within a tolerance of less than $10^{-4}$.} mixing length parameter, namely $\alpha_\mathrm{ML}=2.0$. However, for the comparison we also used a much lower value, namely $\alpha_\mathrm{ML}=1.00$ (low convection efficiency models) which produces cooler models with inflated radii \citep[see e.g.,][]{tognelli2012,tognelli2018}. Evolutionary tracks of mass $\ge$1.20~M$_{\odot}$ have been 
computed adopting a core convective overshooting parameter $\beta_\mathrm{ov}=0.15$, following the recent calibration of the Pisa models by means of the TZ Fornacis eclipsing binary \citep[][]{valle2017}. 

From the models (in the mass range 0.1-20~M$_{sun}$) we obtained isochrones in the age range [0.1, 100]~Myr with a variable spacing in age (down to $\delta t = 0.1$~Myr), allowing for a very good age resolution.
For the reference set of models we used [Fe/H]$= +0.0$, corresponding to an initial helium abundance $Y= 0.274$
 and metallicity $Z=0.013$ given the \citet{asplund2009} solar heavy-element mixture. 

We have converted the theoretical results into the OmegaCAM $r$ and $i$ absolute AB magnitudes using bolometric corrections we computed using a formalism similar to that described in \citet{girardi2002}, employing the MARCS 2008 synthetic spectra library \citep{gustafsson2008}, which is available for $T_\mathrm{eff} \in [2500,\,8000]~$K, $\log g \in [-0.5,\,5.5]$, and [Fe/H] $ \in[-2.0,+0.5]$ completed with the \citet{castelli2003} models for $T_\mathrm{eff} > 8000$~K.

Next we compared the evolutionary models with the data in order to identify the best fitting isochrones for the observed PMS populations. This was performed by using the
 same Bayesian maximum likelihood technique described in \citet{randich2018}. Briefly, the method computes the distance of each star in the observational plane from a given isochrone extracted from our database. The square of the distance is used to define the likelihood of a single star. Then, the total likelihood is defined as the product of all the single star likelihoods.

The parameters to be recovered in this work are the age $t$ and the reddening E(B-V). Thus, along with the grid in age, we also built up a fine grid in E(B-V). To derive the extinction in a given band (i.e., $r$ and $i$) we adopted the \citet{cardelli1989} extinction law by re-computing the extinction coefficients $A_r/A_V$ and $A_i/A_V$ for the OmegaCAM filters employed.  In the following we have performed the parameters recovery under two assumptions: 1) the spread is caused by unresolved binary stars and 2) the spread is actually due to multiple populations.

In the first case, we derived the age and reddening running the recovery over the whole dataset, allowing for the presence of a binary sequence. We have imposed that the binary sequence has a fixed secondary-to-primary mass ratio $q=0.8$ and 1. However, to achieve such a large separation as visible in the CMD, a value of $q=1$ is preferred.

In the second case, that is, in case of multiple populations, we adopted another strategy. First, we selected the members belonging to each population (see Sect~\ref{sec:seq}): the selection identified three populations. We imposed the constraint that all the three populations contained in our sample are affected by the same reddening, in agreement with B17.

Such a condition has been implemented in our recovery code in four   steps. First, we obtained the total likelihood -- which depends on the age and E(B-V) -- for each population. Then, we marginalized over the age to obtain a likelihood which is a function of only E(B-V). This step produces the distribution of the possible E(B-V) for each population. We multiply the E(B-V) distributions obtained at the previous step (for each population), to obtain the E(B-V) distribution for the whole dataset under the assumption that the best E(B-V) is the same. Then, as the best E(B-V) of the whole dataset we chose the maximum of such a distribution. As a final step, we run again the recovery for each population, but using this value of E(B-V) as a prior (Gaussian prior), to derive only the age of each population.

In both cases, to obtain the confidence interval on $t$ and E(B-V), we adopted a Monte Carlo simulation, that is, we perturbed independently each datum within its uncertainty range, to create a given number $N_\mathrm{pert}=100$ of representations of the ONC population. For each representation we derived the age and reddening using the method discussed above, to obtain a sample of ages and reddening. Then, we defined the best value as the mid of the ordered sample (in $t$ and E(B-V)) and the upper and lower extreme of confidence interval as the 84th and 16th percentile of the distribution, respectively.\\
The best fitting parameters are,\\
$\mathrm{age_{pop1}}=4.5^{+1.5}_{-1.2}$ Myr,\\
$\mathrm{age_{pop2}}=2.1^{+0.5}_{-0.4}$ Myr and\\
$\mathrm{age_{pop3}}=1.4^{+0.3}_{-0.2}$ Myr,\\
with E(B-V)=$0.052\pm 0.020$ begin consistent with E(B-V) from B17. The corresponding $\chi^2$ is $\chi^2 = 3.6$. The best fitting isochrones are shown in Fig.~\ref{fig:CMD_sum} (right panel).

\subsection{The apparently old "scattered" objects in the CMD}
After applying selection criteria C1(Eq.~\ref{eq:C1}),1-$\sigma$ C2(Eq.~\ref{eq:C2}) and C3 we expect to be left 
with mainly ONC members which are PMS stars. 
However in right panel of Fig.~\ref{fig:CMD_sum} there is clearly 
a noticeable group of stars that do not appear to be PMS stars, that is, the objects are bluer than the majority of stars. \cite{Manara2013} studied two such objects in detail using broadband, intermediate resolution VLT X-shooter spectra combined with an accurate method to determine the stellar parameters and the related age of the targets. They show that the two  selected stars are actually as young as the bulk of the ONC stars. They conclude that, if only photometry is used as an age estimator then especially for bands sensitive to the presence of accretion, like the here used $r$ and $i$, several accreting objects may appear scattered from the bulk PMS population on the CMD. That is accretors are most likely the explanation for the presence of the blue-wards scattered points in the CMD.  We will investigate these bluer targets in detail in a follow-up work (Beccari et al., in preparation).\\

\section{Apparent binary and triple systems}
\label{sec:Rbin}

In the previous section we investigated  the impact of unresolved binaries on the detection of multiple populations among the PMS of the ONC. We are now interested in exploiting the potential of the Gaia DR2 catalog in combination with the OmegaCAM photometry to investigate resolved multiple systems. The seeing limited OmegaCAM photometry aims at resolving binaries having separations $\gtrapprox 0.7"$. 
While the resolution of Gaia DR2 is 0.4", \cite{Ziegler2018} show that Gaia is only able to reliably recover all binaries down to $1.0^{\arcsec}$ at magnitude contrast as large as 6 magnitudes.

One method to investigate resolved apparent binary candidates is to construct the so-called Elbow plot 
\citep[e.g.,][]{Larson1995,Gladwin1999}. The method consists in calculating the distribution of the number density of observed targets ($\Sigma$) 
as a function of on-the-sky separation ($\theta$). As shown in \citet[][]{Gladwin1999} the presence of an elbow in the density distribution described above is caused by the presence of resolved binaries.
In short,  for each star in our catalog we divide the surrounding area of sky into a set of annuli, by drawing circles of radius $\theta_i$ centered on the star, with $\theta_i=2\theta_{i-1}$ and $i\geq1$. $\theta_0$ is chosen to be well below the smallest separation in the sample. Next we count the number of companion stars in each annulus and we calculate the $\Sigma(\theta)_i$ as follows, 
\begin{eqnarray} 
\Sigma(\theta)_i = \frac{N_i}{\pi N_{\mathrm{tot}} (\theta^2_i-\theta^2_{i-1})}\,,
\end{eqnarray}
where $N_i$ is the number of stars in the $i$th bin, $N_{\mathrm{tot}}$ is the total number of stars and $\theta = (\theta_i+\theta_{i-1})/2$ is the angular separation of measured sources. 

\begin{figure}[ht]
	\scalebox{1.0}{\includegraphics{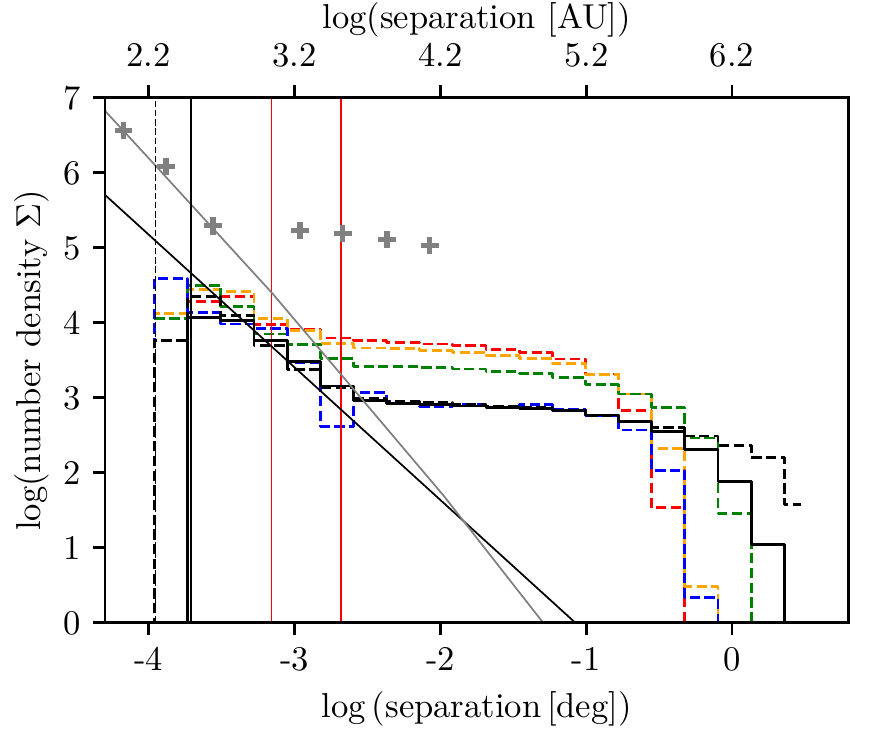}}
	\caption{The Elbow plot showing the surface number density as a function of separation on the sky.
    The black dashed line shows the original Gaia data sample, the black solid line shows the cross-matched, that is, initial, catalog after the C1 and 2-$\sigma$C2 are applied. 
    We fit the elbow part of the diagram obtaining a slope value $ -1.7 \pm 0.05$. The gray data points and the DM slope are from \cite{Simon1997} having a smaller field of view by using the HST. The gray line is the \cite{Duquennoy1991}  distribution of binary stars of G-type primaries that has been reformulated to the number density $\Sigma$ by \cite{Gladwin1999} (their eq. (5)).
   The vertical lines (from left to right) are the resolution limit of Gaia DR2, the resolution limit of OmegaCAM, respectively, 
   the first red line shows ($1\cdot 10^3$ au) as the definition of wide binaries, and the last line shows 
 the start of the Elbow feature ($3\cdot 10^3$ au). 
  The color histograms are constructed for central circular regions (see Fig.~\ref{fig:bin_sky} ) to see how the potential of detection of wide binaries changes with the chosen spatial region. 
  }
	\label{fig:knee}
\end{figure}

We show in Fig.~\ref{fig:knee} the Elbow plot as calculated using the stars sampled with the Gaia DR2 
catalog alone (black dashed line) and Gaia DR2 cross-matched with the OmegaCAM catalog (black solid line). 
We applied the C1 and 2$\sigma$C2 criterion to both catalogs. The C1 and 2$\sigma$C2 criterion reduces significantly the contamination by objects not belonging to the ONC, but on the other 
hand keeps the data sample as complete as possible.
The slope of the elbow is $ -1.70 \pm 0.05$ which suggests incomplete 
data sample toward smaller separations as slope values for complete samples in binary-devoted studies are $\approx -2$ \citep{Gladwin1999}. This is consistent with the recovery completeness found by \cite{Ziegler2018} showing that Gaia DR2 is complete for multiple systems separated by at least 1 arcsec.  \cite{Prosser1994} used the Hubble Space Telescope 
to probe the densest central region of the ONC and \cite{Simon1997} then created 
the elbow plot. The data are plotted as gray crosses in Fig.~\ref{fig:knee}.
The difference between the results of \cite{Simon1997} and our study is most likely
caused by the different field of view, as \cite{Simon1997} looked at the inner-most part of the ONC, as well as by the fact that using Gaia, we are able to successfully remove fore- or background contaminators.

\begin{figure}[ht!]
	\scalebox{1.0}{\includegraphics{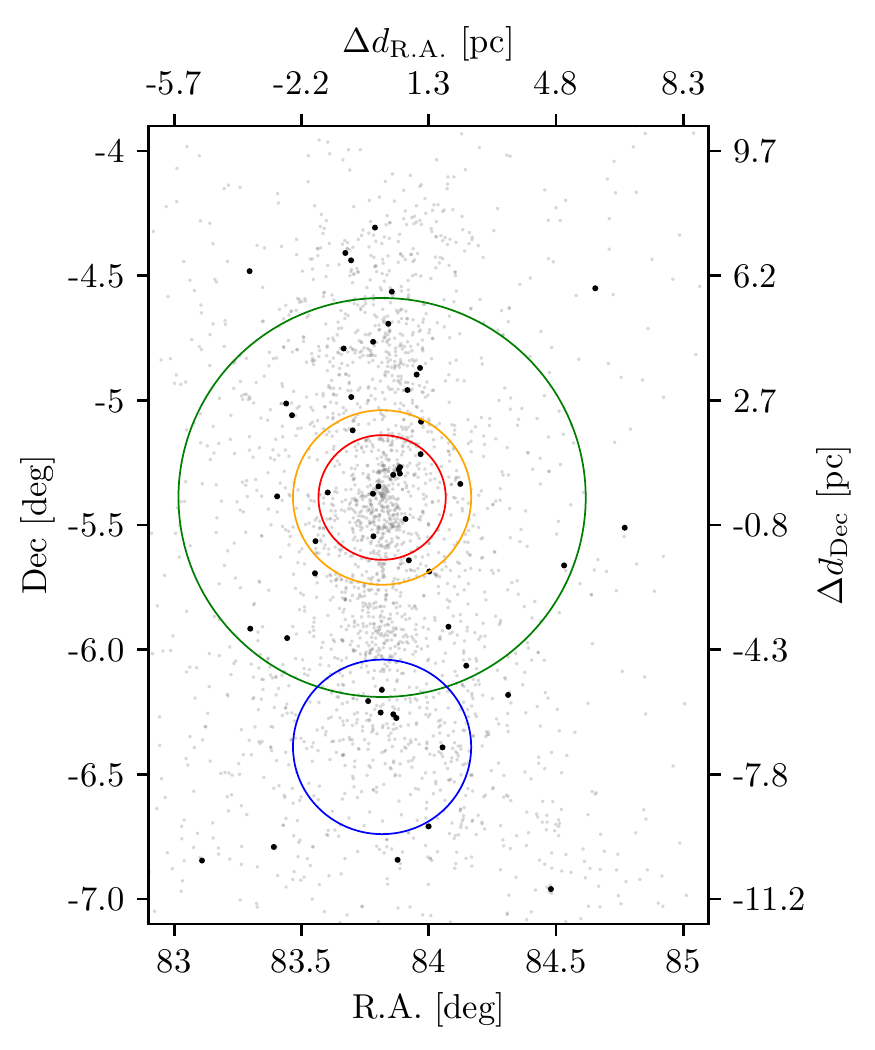}}
	\caption{On-the-sky distribution of the wide (on-the-sky component separation between 1000 and 3000 au) binaries. Only one component of the binary star candidates is plotted. The red circle shows the region, 15 arcmin = 0.25 deg, investigated by \cite{Scaly1999} who found only four   candidates.  The orange (radius of 0.35 deg)  circle shows the region we selected to test how the Elbow plot depends on the investigated spatial scale. The green circle is the (1 deg) large testing region and the blue circle (radius 0.35 deg) is outside where the stellar density is lower.
    }
	\label{fig:bin_sky}
\end{figure}

From the Elbow plot we can see that the overabundance of stars, suggesting the presence of multiple systems, starts at $\approx 3000$ au. Interestingly, \cite{Scaly1999}, based on a proper motion study, suggested that there should be no wide binaries with separations larger than $1000$ au. 
Therefore we decided to investigate the indications of the presence of wide binaries (binaries with apparent separations larger than $1000$ au) in our data in more detail. 

To investigate the effect of crowding on the Elbow plot we divide the sampled region in four   circular regions. Three of the regions are concentric and centered on the ONC center, where the stellar crowding is more severe and hence the completeness expected to be lower. A fourth circle has been centered at 1 deg distance toward the south with respect to the ONC. Such region is located well in the outskirt of the cluster, where the stellar distribution is quite homogeneous and the stellar density very low. We show the location of the circles in the sky in Fig.~\ref{fig:bin_sky} while in Fig.~\ref{fig:knee} we show with the same color code the Elbow plot only for objects within each circle. Clearly, the detection of wide binaries in the central region is challenged by most likely the high stellar density. The more we expand toward the external regions, the detection becomes more and more significant (see the blue line).

Next, we use parallax and proper motion cuts to minimize potential fore- or background contaminants (C1 and C2 but assuming $3 \sigma$ tolerance as here we aim to remain as complete as possible).
There are 86 candidates having apparent separation smaller than $3000$ au as defined by the Elbow plot, and larger than $1000$ au. Such pairs with separations between 1000 and 3000 au are our wide binaries candidates. 

As additional criterion for two stars being a binary candidate we implement the relative proper motion criterion introduced by \cite{Scaly1999}. That is, we check the relative proper motions of all the candidates and require their value to be 0 within the $3 \sigma_{\mu}$ proper motion uncertainty (the average uncertainty, $\sigma_{\mu}$, is larger than the escape velocity from a system with 1000 au separation assuming Solar masses of both components). 
After this we have 60 wide binary candidates from which ten are in the inner 15 arcmin region used by \cite{Scaly1999}. \cite{Scaly1999} found only four candidates in the very same region. All the binary star candidates fulfilling the separation and the pm criteria are listed in Tab.~\ref{tab:bin_sky}, including also binary candidates with separations smaller than 1000 au.

\begin{figure}[ht!]
	\scalebox{1.0}{\includegraphics{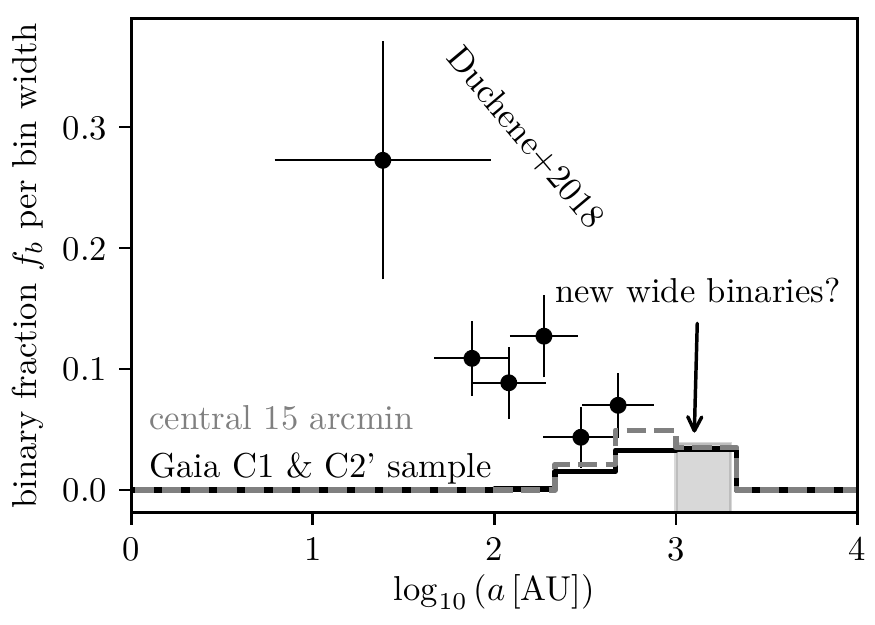}}
	\caption{The plot shows the binary fraction, defined as the number of binaries over the total number of objects, normalized by the bin width. The black points are data for the ONC from \cite{Duchene2018}. The black histogram shows $f_b$ for the spatially not constrained region, that is Gaia DR2 data with C1 and C2 selections. The gray dashed histogram shows the data only for the central part that is the same as in the study by \cite{Scaly1999}. The last bin showing wide binaries in the range 1000-3000 au represents newly discovered potential binaries that seemed to be absent \citep{Scaly1999} (see Fig.~\ref{fig:bin_sky}). The limit 3000 au is motivated by the beginning of the elbow in Fig.~\ref{fig:knee}. 
	}
	\label{fig:bina}
\end{figure}

In Fig.~\ref{fig:bina} we construct the normalized binary fraction distributions defined as the number of binaries over the total number of objects, and compare it with the previous study of \cite{Duchene2018}. 
We extend the separation range of the \cite{Duchene2018} study and find excess of wide binaries in comparison to \cite{Scaly1999}. We compare the central 15 arcmin around the ONC, that is, the same area studied by \cite{Scaly1999}, with the full region studied here and find comparable binary fraction.

The main reason why we are able to detect more of these wide apparent binary star candidates than \cite{Scaly1999} is likely due to the fact that Gaia DR2 allows us, for the first time, to obtain a clean sample of the ONC region only. The contamination of other sources is otherwise too high. We note that due to the crowding at the very center - in the ONC - the detection of the wide pairs are more likely due to chance projections. Our findings are sharpened up by the proper motion study, nevertheless obtaining the 3D velocity information would provide an additional test.

\subsection{Apparent triple systems}
In the Gaia data sample, that is Gaia only catalog with C1 parallax cut and 2-$\sigma$C2 proper motion cut, we identified five apparent triple systems and only one of these has component relative 
proper motions consistent with zero. 
To identify triple system candidates we at first identified the closest potential binary companion candidate and quantified the center of light of the candidate binary. We then find its closest neighbor. If the closest neighbor has a projected separation smaller than 3000 au (the maximal separation captured by our data) then the three neighbors are classified as an apparent triple star candidate.
The identified candidates are summarized in Tab.~\ref{tab:triple}.

\begin{table*}
	\caption{Identified apparent triple system candidates.}
	\label{tab:triple}
	\centering 
	\begin{tabular}{c c c c c c c c}
		\hline\hline 
		ID1 & ID2 & ID3 & $a_1$ [deg.] & $a_2$ [deg.] & $G_1$& $G_2$ & $G_3$\\  
		\hline                                 
		\textcolor{gray}{3017367121346098944} & \textcolor{gray}{3017367121333529984} & \textcolor{gray}{3017367086986361216} & \textcolor{gray}{0.00026} & \textcolor{gray}{0.00197} & \textcolor{gray}{17.2} & \textcolor{gray}{15.5} & \textcolor{gray}{16.8}\\
		\textcolor{gray}{3017244388347833728} & \textcolor{gray}{3017244388350754816} & \textcolor{gray}{3017244388350754944} & \textcolor{gray}{0.00040} & \textcolor{gray}{0.00176} & \textcolor{gray}{14.9} & \textcolor{gray}{16.7} & \textcolor{gray}{15.4} \\
		\textcolor{gray}{3017242532924887936} & \textcolor{gray}{3017242532921992320} & \textcolor{gray}{3017242528629466624} & \textcolor{gray}{0.00042} & \textcolor{gray}{0.00124} & \textcolor{gray}{15.0} & \textcolor{gray}{14.6} & \textcolor{gray}{17.8} \\
		\textcolor{gray}{3209528905960092160} & \textcolor{gray}{3209528905962364160} & \textcolor{gray}{3209528905962363904} & \textcolor{gray}{0.00049} & \textcolor{gray}{0.00111} & \textcolor{gray}{17.5} & \textcolor{gray}{15.4} &\textcolor{gray}{14.3}\\
		3017135399259048704 & 3017135394962860800 & 3017135394965152000 & 0.00052 & 0.00146 & 15.3 &15.2 &16.2\\
		\hline  
	\end{tabular}
	\tablefoot{
		The apparent triple system candidates identified based on the projected distance. Only the bottom-most line, the black one, fulfills the additional constraint of all putative companions having identical proper motions within 3-$\sigma$ uncertainty range. 
	}
\end{table*}

\section{Summary and conclusions} \label{Sec:Conc}
The ONC star forming region has been studied in detail using the data from Gaia DR2 in combination with deep OmegaCAM photometric data. We summarize our main findings bellow: 
Firstly,   
we used parallaxes and proper motions to separate the  ONC stellar population
from background and foreground contamination using Gaia DR2. Cross-matching with the OmegaCAM photometric catalog provides high-quality photometry in $r$ and $i$ filters. This fact turns to be critical in studying the properties of the stellar populations in the ONC as Gaia DR2 photometry does not provide reliable magnitudes in crowded regions (see Appendix~\ref{sec:filt}).
Using these data we are able to construct the CMD and detect two well separated PMSs 
with a suggestion for the existence of a third one, despite Gaia DR2 data being incomplete in the central high density regions where the members of the second and third sequences are mostly located.

To explore the effect of unresolved binary stars we applied filtering as thoroughly as possible. 
The adopted filtering criteria are able to remove basically all unresolved binary stars contributing to the second sequence. In order to probe the capability of our filtering criteria to isolate unresolved binaries, we identified in our catalog 64 unresolved binaries from~\cite{Tobin2009}. We applied our filtering selection to these stars and demonstrate that we are ab;e to safely identify and filter-out 80\% of unresolved binaries. 

We confirm the previous result of B17 that a single stellar populations with binaries (or triples) can not alone describe the observed distribution of stars in the CMD. Our study supports the hypothesis that the PMS population of the ONC is better described as three stellar populations with different ages. We use PMS models calculated using the Pisa stellar evolutionary code to find the best fitting isochrones describing the population on the CMD. We estimate an age difference of $\sim3$ Myr between the youngest ($\sim1.4$Myr) and the oldest ($\sim4.5$Myr) stellar population. In line with B17 the youngest sequences appear to be more concentrated toward the ONC center with respect to the oldest population.
We perform a series of statistical tests with the aim of identifying kinematic differences between the old and young stellar population (see Sect~\ref{sec:seq}). We find a mild indication that the individual sequences might have different parallax distributions with the older population slightly closer to us along the line of sight with respect to the youngest one. While this result is in principle in line with the youngest population being still more embedded in the molecular clouds, we note that the uncertainties in the Gaia DR2 data are still too big to allow us to perform such a detailed study with solid statistical significance. More precise data (or additional data - like radial velocities) will be necessary to confirm these results. 

Interestingly, even adopting a rigorous filtering strategy via accurate parallaxes and proper motions selection criteria, the stars belonging to the ONC seem to occupy a $\approx100$ pc region in the line-of-sight direction and $\approx 20$ pc on the sky. Hence this study well extend behind the canonical size of the ONC star cluster, which is only few a pc large \citep{Hillenbrand1997}. While it is well know that the Orion Nebula is in fact a complex star forming region, hosting several episodes of star formations, we stress here that the precision of the Gaia DR2 parallaxes implies, for a 10\% relative error, a 40 pc uncertainty at the distance of the ONC. 

Secondly, we use our final catalog of bona fide ONC member to study the population of apparent resolved multiple objects in the region. 
We analyze apparent multiple system candidates using the elbow plot. The position of the elbow suggests that wide binaries that we are able to capture have a separation of up to $\approx 3000$ au ($\approx 7$ arcsec). However, this depends on the investigated region and becomes less clear in the dense ONC center. 
Thus we identify all the targets having a separation smaller than 3000 au and identical proper motions within the measurement uncertainty. This allows us to detect in total 91 targets out of which 60 have a separation between 1000 au and 3000 au. 
\cite{Scaly1999} suggested that wide binaries (semi-major axis larger than 1000 au corresponding to an orbital period longer than $28,000\,$yr for a system mass of $1.3\,M_\odot$) are largely absent from the ONC implying that ONC-like clusters cannot be at the origin of the majority of field stars. Here we have found evidence for a large population of wide binaries in the ONC with projected separation up to 3000 au, or an orbital period of $\approx 144,000$~yr for a system mass of $\approx 1 M_{\odot}$. 

Using detailed stellar-dynamical models of the ONC (expanding, collapsing and dynamical equilibrium models) \cite{Kroupa1999} showed that, depending on the dynamical state of the ONC, wide binaries are expected to be present. Moreover, the way binary fraction varies radially can put new constraints on the dynamical state of the ONC. Evidence for the breakup of wide binaries after falling through the cluster has been found in the survey for optical binaries in the ONC by \cite{Reipurth2007}.
Even more realistic N-body models have been computed by 
\cite{Kroupa2001}
who assumed the ONC to be post-gas expulsion, that is, to have just emerged from the embedded phase. These authors also show that the ONC ought to have a population of wide binaries depending on how concentrated the ONC was prior to the onset of the removal of the residual gas (compare the upper and lower panels in their Fig.10). Therefore, if there were no wide binaries, then the pre-gas-expulsion ONC would have corresponded to a model with a half-mass radius close to $0.2\,$pc, while the presence of wide binaries would require it to have been close to $0.45\,$pc. While firm conclusions cannot yet be reached on the dynamical state of the ONC, in particular also because the dynamical state of very young clusters with masses near to that of the ONC may be complex \citep{Kroupa2018, Wang2018}, this discussion shows the importance of surveying for binaries in the ONC and for correlating them with their ages, positions and motion vectors relative to the ONC.

\begin{acknowledgements}
TJ thanks Hans Zinnecker and Pavel Kroupa for interesting and fruitful discussions and useful comments to the manuscript. In addition TJ acknowledges discussions with Josefa Gro{\ss}schedl and Stefan Meingast, who found important formatting error in the Tab. A.1.
We thank Fr\'ed\'eric Arenou for clarification of the photometric excess parameter in the Gaia DR2 catalog. PGPM, ET and SD acknowledge INFN ''iniziativa specifica TAsP''. 
CFM acknowledges an ESO Fellowship. Based on observations collected at the European Southern Observatory under ESO program 098.C-0850(A). 
\end{acknowledgements}

\bibliographystyle{aa}   
\bibliography{library}

\Online

\begin{appendix} 

\section{Comparison of OmegaCAM photometry with Gaia DR2 photometry} \label{sec:filt}
In the following text we demonstrate that additional photometric data set is needed in dense 
region as ONC to be able to study the stellar populations. We follow Gaia DR2 release work by \cite{Arenou2018} 
and show that the photometric information in Gaia Gp and Rp filters gets affected by crowding and is not suitable for precise study.
 \cite{Arenou2018} introduced the photometric 
excess criterion, 
\begin{eqnarray*}
1.0 + 0.015(G_{BP} - G_{RP})^2 < E < 1.3 + 0.06(G_{BP}-G_{RP})^2 \,,
\end{eqnarray*}
where $E$ is 
defined as the flux ratio in the three different Gaia passbands, 
\begin{eqnarray}
E = \frac{I_{BP}+I_{RP}}{I_G}\, .
\end{eqnarray}
The Gaia DR2 presents the photometric set in broad band photometric filters $G$, $G_{BP}$ and $G_{RP}$.
The photometry measurements suffer from significant systematic effects for the faint sources ($G \geq 19$), 
in crowded regions, nearby sources (such as multiple systems) included. 
The G band is measured using the astrometric CCDs with a window of 
typical size of 12x12 pixels (along-scan x across-scan). However it is 60x12 pixels 
for the BP and RP spectral bands. Thus for a single isolated point-source 
the G band flux has a comparable value with the sum of the integrated fluxes from the BP and RP bands as expected from the bands' wavelength coverage. On the other hand, if 
the source is contaminated by a nearby source or it is extended so it is not fully covered by the G band window, then the G flux is not comparable with the sum of BP and RP fluxes \citep{GAIA_DR1}. 
The photometric excess criterion described above can hence be used to check the accuracy of the photometry in the Gaia DR2 catalog \citep{GAIA_DR2}.

\begin{figure*}[ht]
  \scalebox{1.0}{\includegraphics{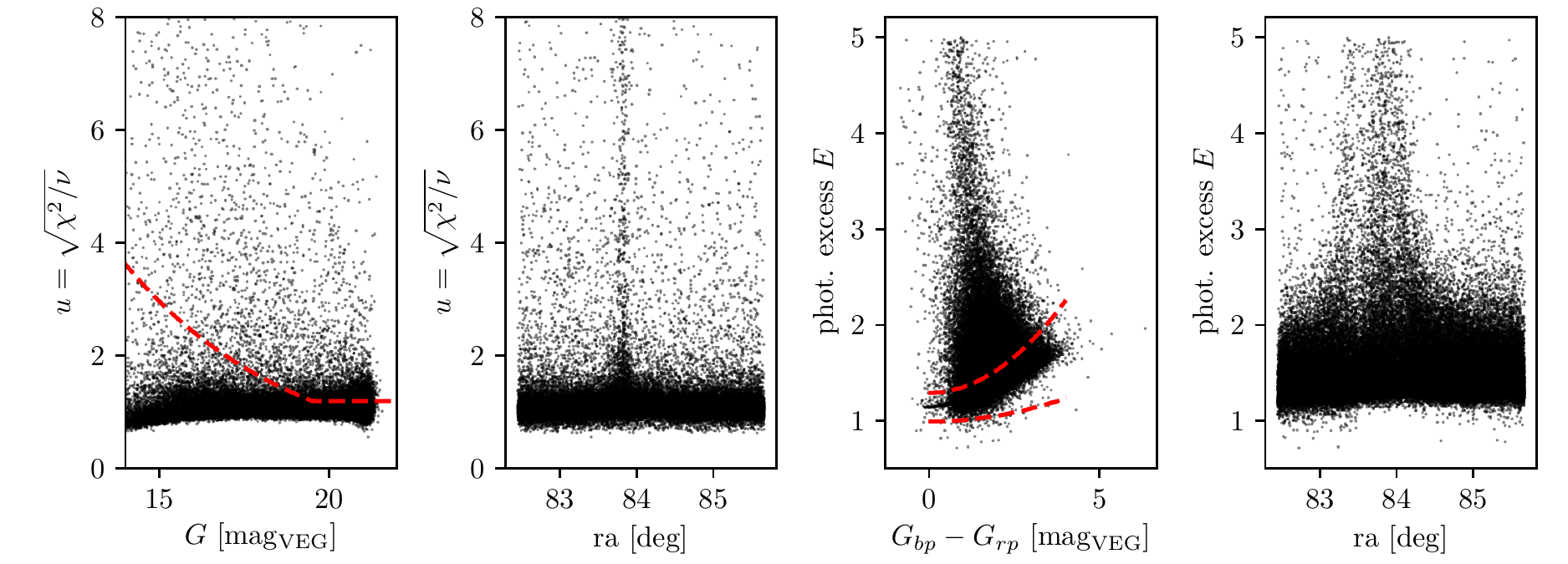}}
  \caption{The black points are the data from the initial catalog. \textbf{First panel:} The astrometric $u$ parameter  as a function 
  of $G$ magnitude. The red dashed line is the $u$ selection criterion from \cite{Arenou2018}. \textbf{Second panel:} shows how this factor depends on R.A. The $u$ parameter is affected by crowding in the central parts of the ONC. 
  \textbf{Third panel:} shows the photometric excess, $E$, parameter as a function of the color in Gaia DR2 filters.
  The region between the dashed red lines shows the $E$ selection criterion  from \cite{Arenou2018}. \textbf{Fourth panel:} shows $E$ as a function of R.A. demonstrating that $E$ and therefore Gaia DR2 photometry in $bp$ and $rp$ filters are significantly affected by crowding.
  }
  \label{fig:uE}
\end{figure*}

We show in Fig.~\ref{fig:uE} (third panel from the left) the distribution of the excess factor $E$ as a function of $G_{BP}-G_{RP}$ color for the entire catalog (gray points) and for the stars selected using the criterion C1 (black points). The red dashed lines indicate the selection limits on this plane as defined using the equation from~ \cite{Arenou2018} and reported above. According to the $E$ limits defined by~\cite{Arenou2018}, only the objects falling in between the two lines should be retained. In the last panel on the right of Fig.~\ref{fig:uE} we show the distribution of $E$ as a function of R.A. Clearly, by filtering the data using such a parameter would  completely exclude the central region of the ONC star forming region where the distribution of $E$ clearly peaks. This is a direct consequence of the fact that the photometric magnitudes $G_{bp}$ and $G_{rp}$ of a given source available in the Gaia DR2 release can be affected by strong contamination by close-by (3.1 x 2.1 arcs$^2$, \citet{Evans2018}) companions in regions affected by high stellar density. We hence decided to not use the $E$ filtering in our analysis and to use the high quality optical photometry available through the OmegaCAM data.

\begin{figure*}[ht]
  \scalebox{1.0}{\includegraphics{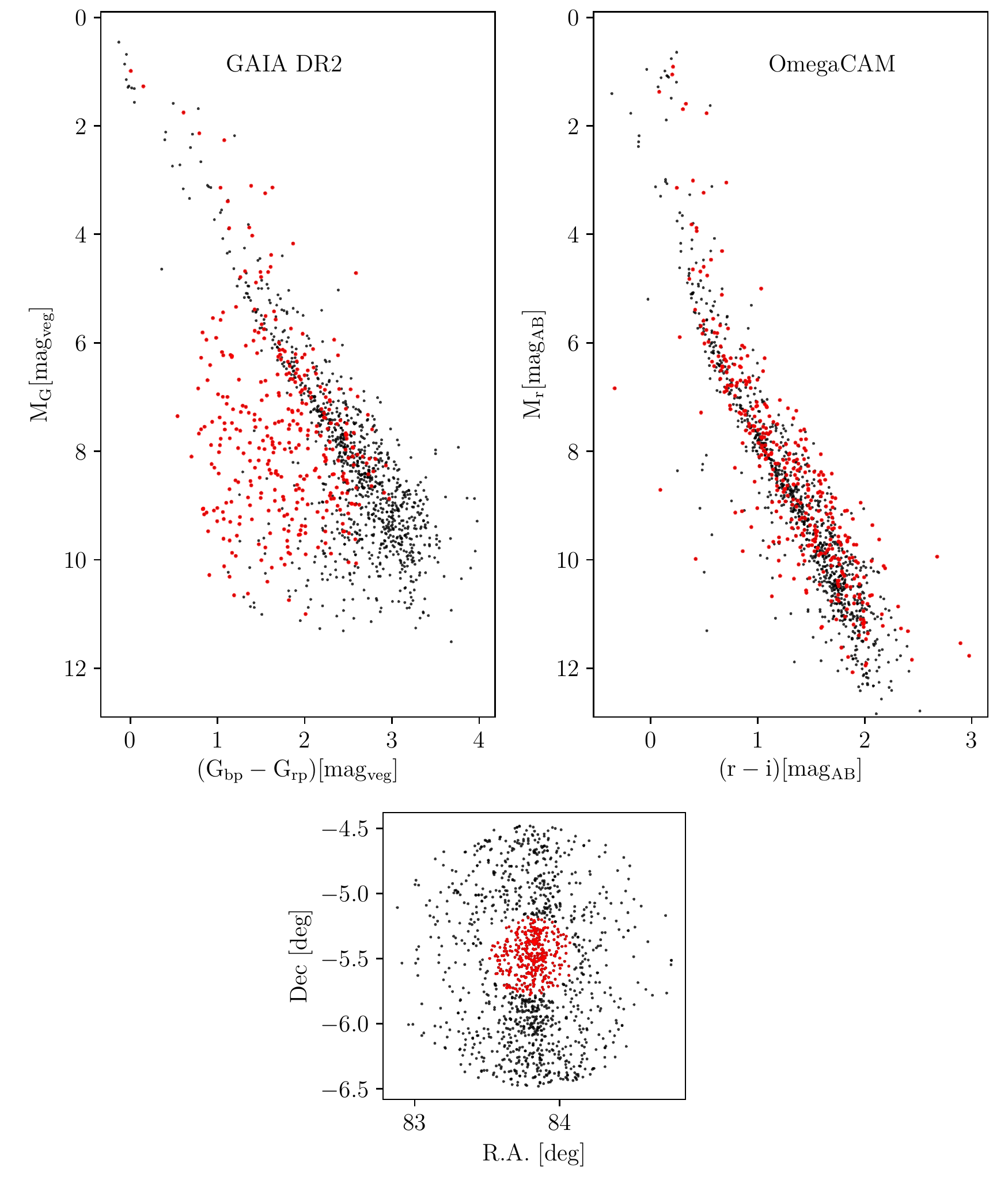}}
  \caption{The black points are the C1 and 2-$\sigma$ C2 selected objects. \textbf{Top panels:} CMD using Gaia photometric filters (left) and 
  using OmegaCAM filters (right). The black points show the one-degree region from the ONC center and the red points the inner 0.2 degree region. Clearly, the Gaia photometry in bp and rp filters is affected by the central crowding. 
  \textbf{Bottom panel:} The R.A. and Dec plot showing the one-degree region and the 0.2 degree central selection (red points).
  }
  \label{fig:crd}
\end{figure*}

To show the effect of crowding on the Gaia photometric filters directly we plot 
the comparison of the CMD for the Gaia photometry and the OmegaCAM one. For 
that we use the C1(eq.(\ref{eq:C1}) selected data sample.
In Fig.~\ref{fig:crd} we select the one degree region around the ONC for which we construct the CMD (black points). We do the same for only the inner region (central 0.2 deg) plotted using red points. We clearly see that the OmegaCAM photometry plots mainly PMS objects with no obvious difference between the black and red data point. 
Contrary to this, Gaia photometry in the bp and rp filters shows significant scatter and deviations from the PMS objects.

\onecolumn

\begin{longtable}{c c c c c c c} 

\caption{Apparent binary stars candidates}\\

number & ID1 & ID2 & $\log_{10}(a)\,$[au] & sep. [arcsec] & G mag1 & G mag2\\
\hline\hline
1 & 3209599721383025792 & 3209599721380315264 & 2.494 & 0.780 & 16.862 & 16.948\\
2 & 3209528523708699392 & 3209528528005247616 & 2.544 & 0.875 & 14.924 & 16.664\\
3 & 3016932397629080448 & 3016932401924820992 & 2.568 & 0.924 & 17.758 & 18.071\\
4 & 3017383201698509312 & 3017383201705272192 & 2.576 & 0.942 & 16.904 & 14.982\\
5 & 3017367121346098944 & 3017367121333529984 & 2.580 & 0.950 & 17.239 & 15.530\\
6 & 3017343310042814464 & 3017343310049098112 & 2.592 & 0.978 & 16.982 & 18.811\\
7 & 3215642671646074240 & 3215642671647475328 & 2.632 & 1.072 & 18.011 & 18.455\\
8 & 3017235901495548928 & 3017235901493475840 & 2.645 & 1.105 & 17.058 & 18.639\\
9 & 3016935867962755968 & 3016935872258388480 & 2.674 & 1.180 & 12.686 & 8.836\\
10 & 3016093543272153856 & 3016093543270824704 & 2.682 & 1.202 & 14.780 & 16.235\\
11 & 3017082244744312576 & 3017082240448693120 & 2.723 & 1.320 & 17.177 & 15.694\\
12 & 3016773827439214720 & 3016773827436894848 & 2.733 & 1.351 & 17.155 & 17.711\\
13 & 3017378803658760960 & 3017378803651996800 & 2.738 & 1.366 & 16.822 & 17.319\\
14 & 3209696061792962432 & 3209696066088278016 & 2.750 & 1.405 & 15.714 & 12.482\\
15 & 3216062272771926144 & 3216062272770760320 & 2.757 & 1.429 & 13.405 & 18.632\\
16 & 3016952502370440064 & 3016952502371760896 & 2.766 & 1.459 & 17.325 & 17.042\\
17 & 3016535684386060544 & 3016535680090105344 & 2.788 & 1.535 & 17.885 & 18.076\\
18 & 3017348257851409536 & 3017348257845029376 & 2.810 & 1.615 & 18.044 & 15.062\\
19 & 3017356602958232064 & 3017356607266117760 & 2.850 & 1.772 & 15.573 & 16.124\\
20 & 3209540313395426176 & 3209540309099093760 & 2.854 & 1.787 & 16.511 & 15.634\\
21 & 3016973290013463936 & 3016973285718581632 & 2.869 & 1.851 & 18.314 & 15.024\\
22 & 3017135399259048704 & 3017135394962860800 & 2.873 & 1.868 & 16.338 & 16.809\\
23 & 3209566667314795264 & 3209566663016374400 & 2.877 & 1.881 & 15.466 & 15.933\\
24 & 3210561587895715328 & 3210561587897894528 & 2.887 & 1.927 & 6.289 & 18.276\\
25 & 3216980983457421440 & 3216980979159662720 & 2.890 & 1.941 & 13.744 & 14.206\\
26 & 3017367293144786304 & 3017367293132202624 & 2.903 & 1.998 & 17.323 & 17.676\\
27 & 3017157595651831168 & 3017157595651831040 & 2.913 & 2.045 & 17.450 & 8.882\\
28 & 3017199548892295808 & 3017199548892295936 & 2.917 & 2.063 & 17.481 & 15.412\\
29 & 3017363582294579456 & 3017363582294579328 & 2.946 & 2.206 & 14.270 & 15.283\\
30 & 3016549492705111936 & 3016549497001135104 & 2.951 & 2.231 & 9.721 & 12.482\\
31 & 3016781356516442624 & 3016781356513940224 & 2.963 & 2.295 & 16.612 & 16.606\\
32 & 3209653597452614144 & 3209653597452613760 & 3.001 & 2.504 & 8.098 & 16.555\\
33 & 3209650883033326848 & 3209650883033326720 & 3.024 & 2.642 & 14.285 & 14.385\\
34 & 3209545845313264256 & 3209545845313264384 & 3.026 & 2.656 & 18.083 & 17.574\\
35 & 3016086740044032512 & 3016086740044032384 & 3.032 & 2.689 & 16.333 & 17.787\\
36 & 3209558386617873920 & 3209558386617873792 & 3.037 & 2.719 & 14.130 & 13.508\\
37 & 3016952326278137344 & 3016952330573066496 & 3.041 & 2.747 & 15.576 & 15.262\\
38 & 3209530276057350656 & 3209530280351892096 & 3.042 & 2.757 & 16.560 & 16.795\\
39 & 3209541412907075456 & 3209541412907075584 & 3.056 & 2.845 & 17.590 & 17.839\\
40 & 3017364303848916992 & 3017364303848916480 & 3.074 & 2.964 & 16.525 & 11.905\\
41 & 3017228273633491712 & 3017228273633491840 & 3.076 & 2.978 & 11.737 & 16.759\\
42 & 3017265244711916928 & 3017265244711916672 & 3.082 & 3.020 & 16.853 & 17.487\\
43 & 3016985281562139136 & 3016985281562139264 & 3.093 & 3.100 & 15.603 & 15.821\\
44 & 3016030493152341632 & 3016030493152341760 & 3.096 & 3.118 & 17.269 & 16.434\\
45 & 3016469060853517824 & 3016469056559465984 & 3.098 & 3.136 & 10.323 & 17.764\\
46 & 3016507303242247552 & 3016507303242247680 & 3.110 & 3.224 & 16.239 & 15.943\\
47 & 3209576528559565184 & 3209576528559728000 & 3.112 & 3.235 & 13.416 & 14.641\\
48 & 3017364441282354176 & 3017364436980916096 & 3.124 & 3.325 & 15.406 & 14.175\\
49 & 3017192608225149568 & 3017192608225149440 & 3.144 & 3.479 & 17.011 & 15.530\\
50 & 3023289881235865216 & 3023289876940144768 & 3.144 & 3.480 & 18.403 & 13.834\\
51 & 3017359562205002880 & 3017347811174480000 & 3.155 & 3.570 & 16.129 & 15.755\\
52 & 3017145707182344320 & 3017145707182344192 & 3.158 & 3.597 & 17.058 & 18.352\\
53 & 3017248786397358080 & 3017248786397531136 & 3.163 & 3.637 & 17.237 & 14.548\\
54 & 3017366468511064960 & 3017366468511064448 & 3.166 & 3.663 & 15.868 & 16.158\\
55 & 3017214869037692800 & 3017214873335520896 & 3.181 & 3.795 & 17.782 & 15.645\\
56 & 3209633290847280256 & 3209633290847279744 & 3.198 & 3.948 & 16.779 & 18.369\\
57 & 3017145638462872576 & 3017147111635948288 & 3.207 & 4.024 & 17.794 & 15.843\\
58 & 3209811652247750272 & 3209811647950503424 & 3.215 & 4.101 & 14.576 & 14.622\\
59 & 3209526844377985152 & 3209526844377984896 & 3.215 & 4.103 & 14.628 & 17.403\\
60 & 3017366640309747072 & 3017366640309746816 & 3.224 & 4.185 & 15.102 & 16.424\\
61 & 3209553576254369792 & 3209553576254369536 & 3.228 & 4.231 & 16.532 & 17.968\\
62 & 3017367391918079744 & 3017367396211298048 & 3.233 & 4.279 & 16.841 & 15.662\\
63 & 3016641370647772672 & 3016641370646103808 & 3.238 & 4.328 & 15.642 & 18.298\\
64 & 3023427522052064896 & 3023427560708065408 & 3.247 & 4.414 & 17.244 & 18.455\\
65 & 3016547710294461440 & 3016547710294461312 & 3.251 & 4.458 & 18.062 & 15.635\\
66 & 3017309981102805760 & 3017309912383329280 & 3.265 & 4.605 & 14.632 & 16.766\\
67 & 3209561371619512704 & 3209561375914929920 & 3.272 & 4.679 & 13.419 & 15.229\\
68 & 3017142752244845056 & 3017142752244845184 & 3.285 & 4.819 & 15.709 & 17.416\\
69 & 3017286479041796480 & 3017286479041796352 & 3.301 & 5.001 & 17.486 & 15.251\\
70 & 3016858975164012928 & 3016858975164013184 & 3.301 & 5.005 & 14.471 & 17.442\\
71 & 3014873940064089472 & 3014873944359395968 & 3.307 & 5.075 & 14.707 & 12.226\\
72 & 3209544024247187584 & 3209544024247187456 & 3.330 & 5.342 & 17.519 & 16.475\\
73 & 3017175462715836416 & 3017175462715713920 & 3.336 & 5.423 & 15.678 & 16.565\\
74 & 3017266962698857216 & 3017266958403468032 & 3.344 & 5.516 & 16.793 & 17.009\\
75 & 3209424795953358976 & 3209424795953358720 & 3.348 & 5.574 & 15.204 & 15.266\\
76 & 3016017264653139840 & 3016017230293401728 & 3.353 & 5.636 & 17.310 & 17.513\\
77 & 3017144538951245184 & 3017144538951245440 & 3.357 & 5.683 & 17.006 & 13.594\\
78 & 3017256757856824448 & 3017256753560684672 & 3.367 & 5.820 & 16.883 & 17.235\\
79 & 3209650608155686272 & 3209650608155686400 & 3.382 & 6.020 & 16.766 & 12.900\\
80 & 3016926045373231744 & 3016926045373231616 & 3.388 & 6.104 & 16.436 & 12.686\\
81 & 3015806124061283968 & 3015806119765504512 & 3.396 & 6.218 & 17.797 & 17.507\\
82 & 3017247515087034240 & 3017247515087033984 & 3.398 & 6.244 & 12.277 & 13.950\\
83 & 3017356435467423616 & 3017356469827161856 & 3.398 & 6.248 & 11.922 & 17.876\\
84 & 3016723803954896256 & 3016723803954896896 & 3.402 & 6.302 & 17.791 & 16.157\\
85 & 3017366537230536832 & 3017366537230535936 & 3.405 & 6.356 & 16.105 & 17.495\\
86 & 3016744037543326592 & 3016744041840721024 & 3.418 & 6.545 & 16.315 & 18.011\\
87 & 3209426582659553408 & 3209426578363441024 & 3.424 & 6.634 & 14.383 & 15.045\\
88 & 3209579449137299072 & 3209579444841918976 & 3.435 & 6.803 & 10.566 & 17.329\\
89 & 3017373203021508736 & 3017373203019748864 & 3.436 & 6.824 & 15.383 & 16.412\\
90 & 3209559864086618240 & 3209559791071547392 & 3.443 & 6.929 & 18.238 & 15.995\\
91 & 3017147661389569536 & 3017147661389568256 & 3.447 & 7.004 & 15.627 & 14.567\\

\footnote{The apparent binary star objects up to 3000 au separation based on the Elbow plot, see Fig.~\ref{fig:knee}. The selection was done on the projected separation and further filtered by the requirement that both components have an identical proper motion within the 3-$\sigma$ uncertainty.}
\label{tab:bin_sky}
\end{longtable}

\end{appendix}

\end{document}